\documentclass[final]{siamltex}
\usepackage{epsfig}

\title{Existence and Stability of Standing Pulses in Neural Networks: I. Existence}

\author{Yixin Guo\thanks{Department of Mathematics, The Ohio State 
University, Columbus, OH 43210  ({\tt yigst@math.ohio-state.edu}).}
        \and Carson C. Chow\thanks{Department of Mathematics,
 University of Pittsburgh, Pittsburgh, PA 15260 ({\tt ccchow@pitt.edu}).}}

\begin{document}

\maketitle

\begin{abstract}
We consider the existence of standing pulse solutions of a neural
network integro-differential equation.  These pulses are bistable with
the zero state and may be an analogue for short term memory in the
brain.  The network consists of a single-layer of neurons synaptically
connected by lateral inhibition.  Our work extends the classic Amari
result by considering a non-saturating gain function.  We consider a
specific connectivity function where the existence conditions for
single-pulses can be reduced to the solution of an algebraic system.
In addition to the two localized pulse solutions found by Amari, we
find that three or more pulses can coexist. We also
show the existence of nonconvex ``dimpled'' pulses and double pulses.
We map out the pulse shapes and maximum firing rates for different
connection weights and gain functions.
\end{abstract}



\pagestyle{myheadings}
\thispagestyle{plain}
\markboth{Yixin Guo and Carson C. Chow}{Existence and Stability of
  Standing Pulses in Neural Networks: I. Existence} 

\section{Introduction}
\label{sec:intro}

The temporary storage of information in the brain for short periods of
time is called working memory~\cite{Baddeley}.  It is known
that the firing activity of 
certain neurons in the cortex are correlated with working memory
states but it is not known what neural mechanisms are responsible for
maintaining the persistent neural
activity~\cite{fuster,Goldmanrakic,Wang}.  Experiments find that a 
specific set of neurons become activated by a memory cue.  They fire
at a rate above their background levels while the memory is 
being held and then return to baseline levels after the memory is extinguished.
When the neurons are active their firing rates are
low compared to their maximal possible rates. Cortical neurons are generally
not intrinsically bistable and do not fire unless given an input that is above a
threshold~\cite{Connors, Mccormick1, Mccormick}.  It has been 
suggested that recurrent excitatory inputs in a network could be 
responsible for maintaining neural activity observed during
memories~\cite{fuster, Goldmanrakic, Grossberg,Mccormick1, 
Mccormick,Wang1,Wang, Wilson,Wilson1,Gutkin, Laing3}.  The
persistent activity is bistable with the background state.  To match
experimental data, a memory network must have the ability to maintain
persistent activity in a selected subset of the neurons while keeping
the firing rates low compared to their possible maximum.

Mathematically, this question has been probed by examining the
existence and stability of localized persistent stationary solutions
of neural network
equations~\cite{Amari77,Coombes,Ermentrout, Ermentrout1, Grossberg,Pinto1, 
Rubin1, Wilson, Wilson1}.  These
localized states have been dubbed `bump
attractors'~\cite{Laing1,Laing2,Laing3,Gutkin,Wang,
  Wilson}.  In a one dimensional network they have also been called
standing 
pulses~\cite{Ermentrout,Pinto1}.  While these simple networks do not
capture all of the biophysical features of cortical circuits they do
capture the qualitative behavior of working memory. 

The coarse-grained averaged activity of a neural network can be
described by~\cite{Amari77,Ermentrout, Grossberg, Wilson, Wilson1}
\begin{eqnarray}
\label{eq:differentiate}
\tau \frac{\partial u(x,t)}{\partial t} = -u(x,t) + \int_\Omega w(x-y)
f[u(y,t)]dy
\end{eqnarray}
where $u(x,t)$ is the synaptic input to neurons located at position $x
\in(-\infty, \infty)$ at time $t \geq 0$, and it represents the level
of excitation or amount of input to a neural element. The connection
function $w(x)$ determines the connections between neurons. The
nonnegative and monotonically non-decreasing gain function $f[u]$,
denotes the \emph{firing rate} at $x$ at time $t$.  We can set the
synaptic decay time $\tau$ to unity without loss of generality.

In his classic work, Amari~\cite{Amari77} considered
(\ref{eq:differentiate}) with a `Mexican Hat' connection function
(i.e.~excitation locally and inhibition distally).  While this is not
biologically realistic for a single layer of neurons, it has been
argued that networks of combined excitatory and inhibitory neurons
with biophysically plausible connections can effectively mimic a
Mexican Hat under certain conditions~\cite{Ermentrout, Kang, Pinto1}.
Amari also made the assumption that $f[u]$ is the Heaviside function.
This approximation made (\ref{eq:differentiate}) 
analytically tractable and he was able to find a host of solutions one
of them being localized stable pulses that are bistable with zero
activity.  Kishimoto and Amari~\cite{Kishimoto} later showed these
solutions also existed for a smooth sigmoidal gain function that saturated
quickly.

Later work considered two populations~\cite{Pinto,Pinto1}, various
connectivity functions~\cite{Coombes, Laing1, Rubin1}, and two
dimensions~\cite{Haskell,Laing3}.  However, all 
used either the Heaviside gain function or a saturating sigmoidal gain
function implying that neurons start to fire when their inputs exceed
threshold and saturate to their maximum rate quickly.  However, in the
brain persistently active neurons fire at rates far below their
saturated maximum~\cite{Colby,Compte,Funahashi, Wang1, Wang}.  
How a network  can maintain 
persistent activity at low firing rates is not fully
understood~\cite{Camperi, Compte, Grossberg, Laing1,Latham1,Latham2, Renart,Wang1, Wang}.

The problem of persistent activity at low firing rates cannot be
addressed with a quickly saturating gain function.  To circumvent this
limitation, we use a nonsaturating piecewise-linear gain function with
a jump i.e. $\beta=0$ (see Fig. \ref{fig:gain_function}) having the form
\begin{eqnarray}
\label{eq:piecewiseligain}
g[u]=\left \{ \begin{array}{ll} \alpha(u-u_{\scriptscriptstyle
T})+\beta& \mbox{\hspace{1.5cm} $u>u_{\scriptscriptstyle T}$} \\ 0 &
\mbox{\hspace{1.5cm} $u \leq u_{\scriptscriptstyle T}$}
\end{array} \right.
\end{eqnarray} 
When the gain $\alpha$ is zero, (\ref{eq:piecewiseligain}) becomes the Heaviside
function scaled by $\beta$. We note that others have considered
piecewise linear gain functions but without the
jump~\cite{Benyishai, Hansel, Seung}.  In these cases,
persistent activity is not possible unless the threshold is set
to zero and the gain to unity where a multi-stable ``line
attractor'' is possible~\cite{Seung}.

In this paper we show the existence of isolated convex standing pulse
solutions (single-pulses) of (\ref{eq:differentiate}).  We consider a
single one dimensional layer of neurons.  Although, this configuration
is a major simplification, it has been shown that such networks
exhibit features present in more realistic architectures.  We
investigate 
how the pulse solutions change when parameters of the gain function and
the connection function change.  We demonstrate the coexistence of two
single-pulse solutions as seen by Amari~\cite{Amari77}, and give
conditions where more than two pulse solutions can coexist.  We also
show the existence of nonconvex ``dimple-pulse'' solutions and
double-pulse solutions.  We derive the stability criteria 
for stable pulses in an accompanying paper~\cite{Guo3}. 

\section{Neural network equations}
\label{sec:connectionandgain}

We study a neural network (\ref{eq:differentiate}) with
lateral-inhibition or Mexican Hat  type connection function $w(x)$ for which excitatory
connections dominate for proximal neurons and inhibitory connections
dominate for distal neurons. In general, $w(x)$ satisfies the
following six properties.

\begin{enumerate}
\label{it:item1}
\item $w(x)$ is symmetric, i.e. $w(-x)=w(x)$,

\item $w(x)>0$ on an interval $(-x_0,x_0)$, and $w(-x_0)=w(x_0)=0$,

\item $w(x)$ is decreasing on $(0,x_0]$;

\item $w(x)<0$ on $(-\infty, -x_0) \cup (x_0, \infty)$;

\item $w(x)$ is continuous on $\Re$, and $w(x)$ is integrable on
$\Re$;

\item $w(x)$ has a unique minimum $x_m$ on $\Re^{+}$ such that
$x_m>x_0$, and $w(x)$ is strictly increasing on $(x_m,\infty)$.
\end{enumerate}

\begin{figure}[!htb]
\centering \epsfig{file=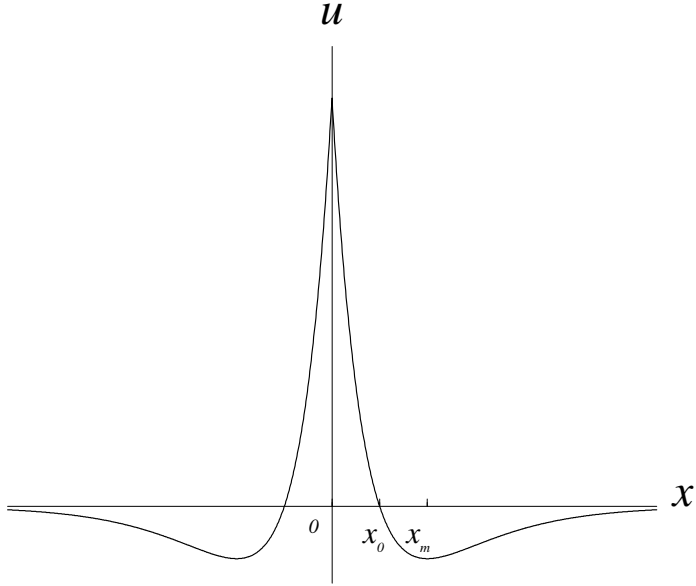, height=3in}
\caption{\small Connection function with $A=2.8,B=1,a=2.6, b=1$.} \centering
\label{fig:Mexi}
\end{figure}

For concreteness, we consider the connection function given by
\begin{equation}
\label{eq:connection}
w(x)=A e^{-a|x|}-e^{-|x|}
\end{equation}
where $a>1$, and $A>1$ guarantee that $w(x)$ obeys properties 1 - 6.
An example of (\ref{eq:connection}) is shown in Fig. \ref{fig:Mexi}.
This connection function is of the lateral inhibition or Mexican Hat
class.  Perhaps, given 
the cusp at zero it should be called a ``Wizard Hat'' function.

For connection function (\ref{eq:connection}),
${\displaystyle x_0=\frac{\ln A}{a-1}}$ and ${\displaystyle
x_m=\frac{\ln aA}{a-1}}.$ The area of $w(x)$ above and below the
$x$-axis represents the net excitation and inhibition in the network
respectively. The total area of (\ref{eq:connection}) is
$2(\frac{A}{a}-1)$. The amount of excitation and inhibition depends on
the ratio of $A$ to $a$. If $A>a$, i.e. $2(\frac{A}{a}-1)>0$,
excitation dominates in the network and if $2(\frac{A}{a}-1)<0,$
inhibition dominates. In the balanced case, $A=a$,
i.e. $2(\frac{A}{a}-1)=0$.

\begin{figure}[!htb]
\epsfig{figure=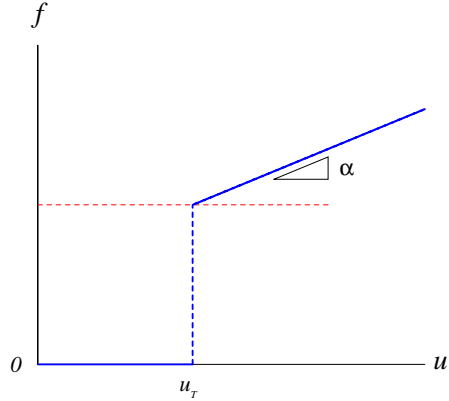,height=2in} \centering
\caption{\small Piecewise-linear gain function.} \centering
\label{fig:gain_function}
\end{figure}
The gain function (\ref{eq:piecewiseligain}) can be written as
\begin{eqnarray}
f[u]=[\alpha(u-u_{\scriptscriptstyle T})+\beta]
\Theta(u-u_{\scriptscriptstyle T})
\label{eq:firingrate}
\end{eqnarray}
where $\Theta(u-u_{\scriptscriptstyle T})$ is the Heaviside function
such that
\begin{equation}
 \Theta(u-u_{\scriptscriptstyle T})=\left\{ \begin{array}{ll} 1 & {\rm
if} \ u > u_{\scriptscriptstyle T} \\ 0 & \mbox{otherwise}
\end{array}\right..
\end{equation}
The gain function (\ref{eq:firingrate}) does not saturate with a
positive slope $\alpha$.  Without loss of generality, we set
$\beta=1$. The gain function (\ref{eq:firingrate}) turns into the
Heaviside function when $\alpha=0$ ( See
Fig.~\ref{fig:gain_function}).


A stationary solution of (\ref{eq:differentiate}) satisfies the
equilibrium equation 
\begin{equation}
\label{eq:integral_equation2}
u(x)=\int_{-\infty}^{\infty}w(x-y)f[u(y)]dy.
\end{equation}
An example of a working memory state can be seen by
considering constant solutions of  (\ref{eq:integral_equation2}).  For
$u(x)=u^0$, the integral equation becomes 
\begin{eqnarray}
u^0=f[u^0]\int_{-\infty}^{\infty}w(y)dy 
\end{eqnarray}
Using (\ref{eq:connection}), the constant solution satisfies
\begin{eqnarray}
\label{eq:u0sol}
u^0=w^0 f[u^0].
\end{eqnarray}
where $w^0=2(A/a-1)$.  From (\ref{eq:u0sol}), we immediately see that
$u^0=0$ is a solution.  In fact, zero is a solution of
(\ref{eq:integral_equation2}) for any positive threshold
$u_{\scriptscriptstyle T}$ and any values of parameters
$a$, $A$, and $\alpha.$
 
\begin{figure}[htb!]
\centering \epsfig{file=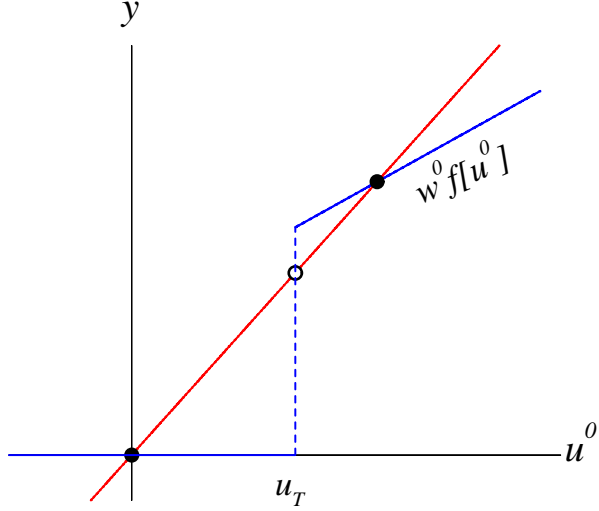, height=2.6in}
\caption{\small Bistability of constant solutions. The solid circles are the
two stable constant solutions and the open circle is an unstable
solution. $w^0$ is the integral of $w(x)$ on its 
domain.} \centering
\label{fig:consdomainm1}
\end{figure}

Inserting gain function (\ref{eq:firingrate}) into
(\ref{eq:u0sol}) gives
\begin{equation}
u^0=w^0(\alpha (u^0 - u_{\scriptscriptstyle T}) +1)
\label{eq:u02}
\end{equation}
The existence of constant solutions can be deduced graphically (see
Fig.~\ref{fig:consdomainm1}).  Nontrivial constant solutions ($u^0>0$)
require $w^0>0$ which means that $A/a>1$.  Thus only for net
excitatory connections are nontrivial constant solutions possible. A
simple stability calculation shows that $\alpha<1$ is necessary for
stability.  Condition (\ref{eq:u02}) shows that for $u_T<0$ and $\alpha<1$,
there is a single stable solution.  If $u_T>0$ and $\alpha>0$ there
can be three solutions (see Fig.~\ref{fig:consdomainm1}).  Two of the
solutions, $u^0=0$ and $u^0>u_T$ are stable.  The third solution at
$u^0=u_T$ is unstable.  For this parameter set, the network exhibits
working memory-like behavior.  The 
network is stable in the background state $u^0=0$.  A transient input
from a memory cue can switch the network into the stable $u^0>u_T$
state which represents the memory.  This is a state of persistent
activity that is sustained by positive feedback.  The state can be
switched off to zero by another transient input when it is no longer
needed.  The next section will examine 
spatially localized pulses that have the same memory property.

\section{Single-pulse solutions}
\label{sec:singlepulse} 
We prove the existence and determine the properties of localized
stationary persistent states which we call single-pulses.  We consider
single-pulse solutions of (\ref{eq:integral_equation2}) that satisfy
the following:
\begin{definition}
\label{de:single-pulse}
{\bf{Single-pulse solution}:}
\label{def:single-pulse}
$$u(x) \left \{ \begin{array}{ll} >u_{\scriptscriptstyle T} &
  \mbox{$\mathrm{if}$ $x\in(-x_{\scriptscriptstyle
      T},x_{\scriptscriptstyle T})$, $x_{\scriptscriptstyle T}>0$}\\
  =u_{\scriptscriptstyle T} & \mbox{$\mathrm{if}$
    $x=-x_{\scriptscriptstyle T},x=x_{\scriptscriptstyle T}$} \\
  <u_{\scriptscriptstyle T} & \mbox{$\mathrm {otherwise}$}
\end{array} \right.$$ such that
${\displaystyle (u,u',u'',u''') \rightarrow (0,0,0,0)}$
exponentially fast as $x \rightarrow \pm \infty$ and $u, u' \in
L^1(\Re)$.  $u$ and $u'$ are bounded and continuous on $\Re$. $u''$,
$u'''$, and $u''''$ are continuous everywhere except at $x=\pm
x_{\scriptscriptstyle T}$ and bounded everywhere on $\Re$. $u(x)$ is
symmetric with $u''(0)<0$; $u(0)$ is the maximum between
$-x_{\scriptscriptstyle T}$ and $x_{\scriptscriptstyle T}$ ( $\mathrm
{Fig.~\ref{fig:bumppic}}.)$
\end{definition}

We note that there also exist pulses where $u''(0)>0$, which implies
$u(0)$ is no longer the maximum of the pulse. We call this solution a
{\em dimple-pulse}. The theorem below gives a range for which there is
no single-pulse solution.

\begin{theorem}
\label{th:nobump}
For fixed $a$, $A$ and $\beta=1$, there is no single-pulse solution if
both ${\displaystyle \alpha<\frac{a}{2A}}$ and ${\displaystyle
u_{\scriptscriptstyle T}>\frac{2A}{a}}$ are true.
\end{theorem}
\begin{proof}
Substituting the
exponential connection function (\ref{eq:connection}) and gain
function (\ref{eq:firingrate}) into the integral equation
(\ref{eq:integral_equation2}) gives
\begin{equation}
u(x)=\int_{\-\infty}^{\infty}(Ae^{-a|x-y|}-e^{-|x-y|})[\alpha(u(y)-u_{\scriptscriptstyle
T})+ 1]\Theta(u-u_{\scriptscriptstyle T})dy.
\end{equation}
Suppose there is a single-pulse solution as defined above when both
${\displaystyle \alpha<\frac{a}{2A}}$ and ${\displaystyle
u_{\scriptscriptstyle T}>\frac{2A}{a}}$ are satisfied. For a single pulse to exist
\begin{eqnarray}
u(0) & = &
\int_{-\infty}^{\infty}(Ae^{-a|y|}-e^{-|y|})[\alpha(u(y)-u_{\scriptscriptstyle
T})+ 1]\Theta(u-u_{\scriptscriptstyle T})dy \nonumber \\ & = &
\int_{-x_{\scriptscriptstyle T}}^{x_{\scriptscriptstyle
T}}(Ae^{-a|y|}-e^{-|y|})[\alpha(u(y)-u_{\scriptscriptstyle T})+ 1]dy
\nonumber \\ & \leq & \int_{-x_{\scriptscriptstyle
T}}^{x_{\scriptscriptstyle
T}}Ae^{-a|y|}[\alpha(u(y)-u_{\scriptscriptstyle T})+ 1]dy \nonumber
\label{eq:eqn1}
\end{eqnarray}
where $u(x)\ge u_T$ is continuous on $I:=[-x_{\scriptscriptstyle T},
x_{\scriptscriptstyle T}]$. $Ae^{-a|y|}$ is integrable on $I$ and
$Ae^{-a|y|} \geq 0$ for all $x \in I$.  By the Mean Value Theorem for
Integrals, $\exists$ $ c^0 \in I$ s.t.
\begin{eqnarray}
u(0) & \le & (\alpha u(c^0) -\alpha u_T+1)
\int_{-x_{\scriptscriptstyle T}}^{x_{\scriptscriptstyle
T}}Ae^{-a|y|}dy \nonumber \\ & \leq & \alpha P u(c^0) + (1-\alpha
u_{\scriptscriptstyle T}) P
\end{eqnarray}
where ${\displaystyle
P=\int_{-\infty}^{\infty}Ae^{-a|y|}dy=\frac{2A}{a}}$.  If $\alpha P
<1$ and $(1-\alpha u_{\scriptscriptstyle T}) \leq 0$ are both true,
then $u(0) < u(c^0)$, $c \in I$.  However, this cannot be true because
$u(0)$ is the maximum of $u(x)$ on $\Re$.  From $\alpha P <1$, we get
${\displaystyle \alpha <\frac{a}{2A}}$. From $(1-\alpha
u_{\scriptscriptstyle T}) \leq 0$, ${\displaystyle
u_{\scriptscriptstyle T} \geq
\frac{1}{\alpha}>\frac{2A}{a}}$. Therefore, there is no single-pulse
when both ${\displaystyle \alpha <\frac{a}{2A}}$ and ${\displaystyle
u_{\scriptscriptstyle T} > \frac{2A}{a}}$ are both
true. In other words if the gain is too low or the threshold too high,
there cannot be a single-pulse. \qquad
\end{proof}

\subsection{Strategy to construct a single-pulse solution}
\label{sec:strategy}
The general approach to studying integral equation (\ref{eq:integral_equation2}) is to
derive an associated differential equation whose solutions are also solutions of the
integral equation. 
We derive the differential equation by using the Fourier transform
\begin{eqnarray*}
F[g(x)]=\int_{-\infty}^{\infty}g(x)e^{isx}dx
\end{eqnarray*}
where $g \in L^{1}(\Re)$ and $s \in \Re$,
with the inverse Fourier transform
\begin{eqnarray*}
g(x)=\frac{1}{2\pi}\int_{-\infty}^{\infty}F[g(x)]e^{-isx}ds.
\end{eqnarray*}

For our conditions on $u(x)$ and $w(x)$, an application of the Fourier
transform to (\ref{eq:integral_equation2}) is well-defined and turns
the convolution into a point-wise product
\begin{equation}
\label{eq:convol}
F[u]=F[w]F[f[u]].
\end{equation}
Computing $F[w]$ in (\ref{eq:convol}) gives
\begin{equation}
\label{eq:equ1}
F[u] =\frac{(2aA+2as^2A-2a^2-2s^2)}{(a^2+a^2s^2+s^2+s^4)}F[f].
\end{equation}
Multiplying both sides of (\ref{eq:equ1}) by the denominator of the right
hand side and using the linear property of the Fourier Transform with
the identities $F[u'']=-s^2F[u]$ and $F[u^{''''}]=s^4F[u]$ gives
\begin{equation}
\label{eq:ode1}
F[u^{''''}-(a^2+1)u''+a^2u] = F[2(aA-a^2)f]+2(aA-1)F[s^2f].
\end{equation}
By the definitions of $u(x)$ and $f[u]$,
$$F[u^{''''}-(a^2+1)u''+a^2u]$$ and $$ F[2(aA-a^2)f]$$ are in
$L^1(\Re)$.

Integrating $F[s^2f]$ by parts yields
\begin{eqnarray*}
\lefteqn{F[s^2f] } \\ &=&\int_{-\infty}^{\infty}s^2e^{isx}f[u(x)]dx \\
& = & \int_{-x_{\scriptscriptstyle T}}^{x_{\scriptscriptstyle
T}}s^2e^{isx}f[u(x)]dx\\ &=&f[u(x_{\scriptscriptstyle
T})](-ise^{isx_{\scriptscriptstyle T}}+ise^{-isx_{\scriptscriptstyle
T}})+f'[u(x_{\scriptscriptstyle T}^-)]u'(x_{\scriptscriptstyle
T})(e^{isx_{\scriptscriptstyle T}}+e^{-isx_{\scriptscriptstyle T}})\\
& & -\int_{-x_{\scriptscriptstyle T}}^{x_{\scriptscriptstyle
T}}e^{isx}\frac{d^2f[u(x)]}{dx^2}dx.
\end{eqnarray*}
Note that $f[u(x)] = 0$ outside of $(-x_{\scriptscriptstyle T},
x_{\scriptscriptstyle T})$ and $F[s^2f] \in L^1(\Re)$.

Applying the inverse Fourier transform to (\ref{eq:ode1}) gives a
fourth order ordinary differential equation
\begin{eqnarray}
\label{eq:general_ode} 
&& u^{''''}-(a^2+1)u''+a^2u = 2(aA-a^2)f[u(x)]+
\\ & & 2(aA-1)\left\{f[u(x_{\scriptscriptstyle
T})]\Delta'(x)+f'[u(x_{\scriptscriptstyle
T}^-)]u'(x_{\scriptscriptstyle T}) \Delta(x)- \frac{d^2f[u(x)]}{dx^2}
\right\} \nonumber
\end{eqnarray}
where
$$\Delta'(x)=\delta'(x-x_{\scriptscriptstyle
T})+\delta'(x+x_{\scriptscriptstyle T})$$
and
$$\Delta(x)=\delta(x-x_{\scriptscriptstyle
T})+\delta(x+x_{\scriptscriptstyle T}).$$ 
Here $\delta(x)$ and
$\delta'(x)$ are defined as~\cite{Folland}
$${\displaystyle \delta(x)=\int_{-\infty}^{\infty} e^{isx}dx},
\mbox{\hspace{1cm}} {\displaystyle
\delta'(x)=is\int_{-\infty}^{\infty} e^{isx}dx}.$$ If $u(x)$ is a
solution of (\ref{eq:general_ode}) where
(\ref{eq:convol})-(\ref{eq:ode1}) hold, then $u(x)$ is also a solution
of (\ref{eq:integral_equation2}).

\begin{figure}[!htb]
\epsfig{figure=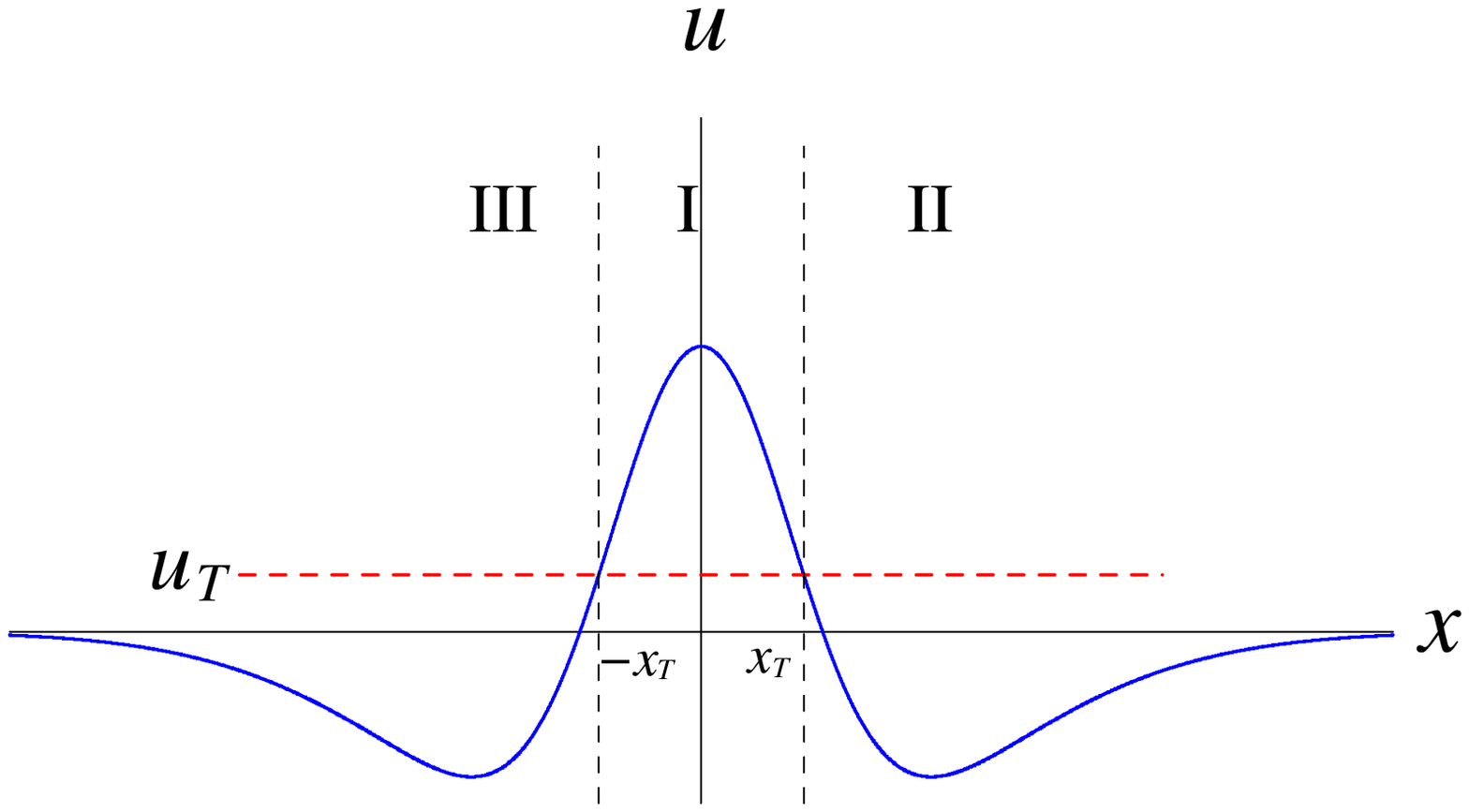,height=2in} \centering
\caption{\small Single-pulse solution.} \centering
\label{fig:bumppic}
\end{figure}

We construct a single-pulse solution as in Fig.~\ref{fig:bumppic} by
decomposing ODE (\ref{eq:general_ode}) into two linear differential
equations:
\begin{eqnarray}
\label{eq:regIODE1}
\mbox{} \hspace{0.7cm} u^{''''}-(a^2+1)u''+a^2u & =& 2a(A-a)f(u)-2(aA-1)
\frac{d^2f[u]}{dx^2},  \mbox{if $u>u_{\scriptscriptstyle T}$ (region
I)} \\
\mbox{}  \hspace{0.7cm} u^{''''}-(a^2+1)u''+a^2u & = &0, 
\mbox{\hspace{3cm} if $u<u_{\scriptscriptstyle T}$ (region II and III)}
\label{eq:regIIODE2}
\end{eqnarray}

We label the solution of (\ref{eq:general_ode}) on regions I, II and III,
$u_{\mathrm I}(x)$ by $u_{\mathrm {II}}(x)$ and $u_{\mathrm {III}}(x)$
respectively.  The solutions $u_{\mathrm I}(x)$, $u_{\mathrm {II}}(x)$ and
$u_{\mathrm {III}}(x)$ must be connected together at
$-x_{\scriptscriptstyle T}$ and $x_{\scriptscriptstyle T}$ to get a
continuous and smooth $u(x)$ on $\Re$.  $u_{\mathrm I}(x)$
and $u_{\mathrm {II}}(x)$ are connected at $x_{\scriptscriptstyle T}$ with five
matching conditions:
\begin{eqnarray}
\label{eq:matchingcondition1}
u_{\mathrm I}(x_{\scriptscriptstyle T}) & = & u_{\scriptscriptstyle T}
\\
\label{eq:matchingcondition2}
u_{\mathrm {II}}(x_{\scriptscriptstyle T})& = & u_{\scriptscriptstyle
T} \\
\label{eq:matchingcondition3}
u_{\mathrm I}'(x_{\scriptscriptstyle T})& = & u_{\mathrm
{II}}'(x_{\scriptscriptstyle T}) \\
\label{eq:matchingcondition4} u_{\mathrm I}''(x_{\scriptscriptstyle T})& = &
u_{\mathrm {II}}''(x_{\scriptscriptstyle
T})-2(aA-1)f(u(x_{\scriptscriptstyle T})) \\
\label{eq:matchingcondition5} u_{\mathrm I}'''(x_{\scriptscriptstyle T})& = &
u_{\mathrm {II}}'''(x_{\scriptscriptstyle
T})-2(aA-1)f'(u(x_{\scriptscriptstyle T}))u'(x_{\scriptscriptstyle T})
\end{eqnarray}
Conditions (\ref{eq:matchingcondition1})-(\ref{eq:matchingcondition3}) are given
by the continuity of $u(x)$ and $u'(x)$.
(\ref{eq:matchingcondition5}) is obtained by integrating
(\ref{eq:general_ode}) over a small neighborhood of
$x_{\scriptscriptstyle T}$. (\ref{eq:matchingcondition4}) is obtained
by integrating (\ref{eq:general_ode}) twice, first with respect to
$x$, then over a small neighborhood of $x_{\scriptscriptstyle T}$.
$u''(x)$ and $u'''(x)$ are discontinuous at $x_{\scriptscriptstyle
T}$, $i.e.$ there are jumps in $u''(x_{\scriptscriptstyle T})$ and
$u'''(x_{\scriptscriptstyle T})$.  Since $u(x)$ is symmetric, similar
matching conditions apply to $u_{\mathrm I}(x)$ and $u_{\mathrm
{III}}(x)$ at $-x_{\scriptscriptstyle T}$.

In region II, the solution for a single pulse that satisfies the boundary conditions is
\begin{eqnarray}
\label{eqa:solregII}
u_{\mathrm {II}}(x)=Ee^{-ax}+Fe^{-x}\mbox{\hspace{3cm}}
\mbox{\hspace{2cm}}E, F \in \Re.
\end{eqnarray}
By symmetry, the solution in region III is
\begin{eqnarray}
\label{eqa:solregIII}
u_{\mathrm {III}}(x)=Ee^{ax}+Fe^{x},\mbox{\hspace{3cm}}
\mbox{\hspace{2cm}}E, F \in \Re.
\end{eqnarray}

In region I, substituting $f[u(x)]=\alpha(u-u_{\scriptscriptstyle
T})+1$
and ${\displaystyle \frac{d^2f[u(x)]}{dx^2}=\alpha u''(x)}$ into
(\ref{eq:regIODE1}) gives
\begin{eqnarray}
\label{eq:regIODE} \hspace{1.cm}u^{''''}-(a^2+1-2\alpha(aA-1))u''+(a^2-2a\alpha(A-a))u
=2a(A-a)(1-\alpha u_{\scriptscriptstyle T})
\end{eqnarray}
The eigenvalues of (\ref{eq:regIODE}) are $\omega_1$, $-\omega_1$,
$\omega_2$, $- \omega_2$ where
\begin{eqnarray}
\omega_1^2 & = & R+S \label{eiv1}\\ \omega_2^2 & = & R-S \label{eiv2}
\end{eqnarray}
with
\begin{equation}
R=\frac{(a^2+1-2\alpha(aA-1))}{2},
\label{R}
\end{equation}
\begin{equation}
S=\frac{\sqrt{\Delta}}{2},
\label{S}
\end{equation}
and
\begin{equation}
\Delta=(a^2+1-2\alpha(aA-1))^2-4(a^2-2a\alpha(A-a)).
\label{Delta}
\end{equation}
Imposing symmetry and $u'(0)=0$, the general solution of
ODE(\ref{eq:regIODE}) can be written in the form:
\begin{equation}
\label{eq:generalsolregI}
u_{\mathrm I}(x)=C(e^{\omega_1x}+e^{-\omega_1x})+D(e^{\omega_2x}+e^{-\omega_2x})+U_0
\end{equation}
where
$$
U_0 = \frac{2(A-a)(\beta-\alpha u_{\scriptscriptstyle T})}{a-2\alpha(A-a)}
$$
for $x \in (-x_{\scriptscriptstyle T}, x_{\scriptscriptstyle
T})$, $x_{\scriptscriptstyle T} \in \Re$, $C, D \in \mathbf{C}$, and
 $u_{\mathrm I}(x)\in \Re$.

The single-pulse solutions of (\ref{eq:general_ode}) are found by
matching $u_I$, $u_{II}$, and $u_{III}$ across $x_{\scriptscriptstyle
  T}$ and $-x_{\scriptscriptstyle T}$ using 
the matching conditions (\ref{eq:matchingcondition1}) -
(\ref{eq:matchingcondition2}).   We investigate the existence and
shape of single-pulse solutions as we change the gain and connection
function. 
For simplicity, we call $x_{\scriptscriptstyle T}$ the width of
a pulse although it is actually the half width.  The height of a
single-pulse is $u(0)$.  The firing rate of the pulse is given by $f[u]$.

\subsection{Solutions for the Amari case ($\alpha=0$)}
\label{sec:amari}
Amari found conditions for which single-pulse solutions exist for
(\ref{eq:integral_equation2}) with general Mexican Hat connectivity
and the Heaviside gain function~\cite{Amari77}.  Here, we revisit the
Amari case for the exponential connection function (\ref{eq:connection}).  When
$\alpha=0$ and $\beta=1$, the gain function (\ref{eq:firingrate})
becomes the Heaviside function $\Theta(u)$ and the term ${\displaystyle
2(aA-1)\frac{d^2f[u]}{dx^2}}$ does not exist in ODE
(\ref{eq:regIODE1}).  The eigenvalues (\ref{eiv1}) and (\ref{eiv2})
become simple and the solutions  (\ref{eq:generalsolregI}) and (\ref{eqa:solregII}) are 
\begin{eqnarray}
\label{eq:amsolregI} u_{\mathrm I}(x)& =
&C(e^{ax}+e^{-ax})+D(e^{x}+e^{-x})+U_0, \\ \label{eq:amsolregII}
u_{\mathrm {II}}(x) & = & Ee^{-ax}+Fe^{-x},
\end{eqnarray}
respectively. Applying conditions
(\ref{eq:matchingcondition1})-(\ref{eq:matchingcondition5}) to
(\ref{eq:amsolregI}) and (\ref{eq:amsolregII}) yields the  system 
\begin{small}
\begin{equation}
\label{eq:am1}
Ee^{-ax_{\scriptscriptstyle T}}+Fe^{-x_{\scriptscriptstyle T}} =
u_{\scriptscriptstyle T} 
\end{equation}
\begin{equation}
\label{eq:am2} C(e^{ax_{\scriptscriptstyle T}}+e^{-ax_{\scriptscriptstyle
T}})+D(e^{x_{\scriptscriptstyle T}}+e^{-x_{\scriptscriptstyle
T}})+\frac{2(A-a)\beta}{a} = u_{\scriptscriptstyle T} 
\end{equation}
\begin{eqnarray}
\label{eq:am3} aC(e^{ax_{\scriptscriptstyle
T}}-e^{-ax_{\scriptscriptstyle T}})+D(e^{x_{\scriptscriptstyle
T}}-e^{-x_{\scriptscriptstyle T}}) & =& -aEe^{-ax_{\scriptscriptstyle
T}}-Fe^{-x_{\scriptscriptstyle T}} \\ \label{eq:am4}
a^2C(e^{ax_{\scriptscriptstyle T}}+e^{-ax_{\scriptscriptstyle
T}})+D(e^{x_{\scriptscriptstyle T}}+e^{-x_{\scriptscriptstyle T}}) &=&
a^2Ee^{-ax_{\scriptscriptstyle T}}+Fe^{-x_{\scriptscriptstyle T}} -2
(a A-1) \beta \\ \label{eq:am5} a^3C(e^{ax_{\scriptscriptstyle
T}}-e^{-ax_{\scriptscriptstyle T}})+D(e^{x_{\scriptscriptstyle
T}}-e^{-x_{\scriptscriptstyle T}}) &=& -a^3Ee^{-ax_{\scriptscriptstyle
T}}-Fe^{-x_{\scriptscriptstyle T}}
\end{eqnarray}
\end{small}
\begin{figure}[!htb]
\begin{minipage}{2.5in}
\centering \epsfig{figure=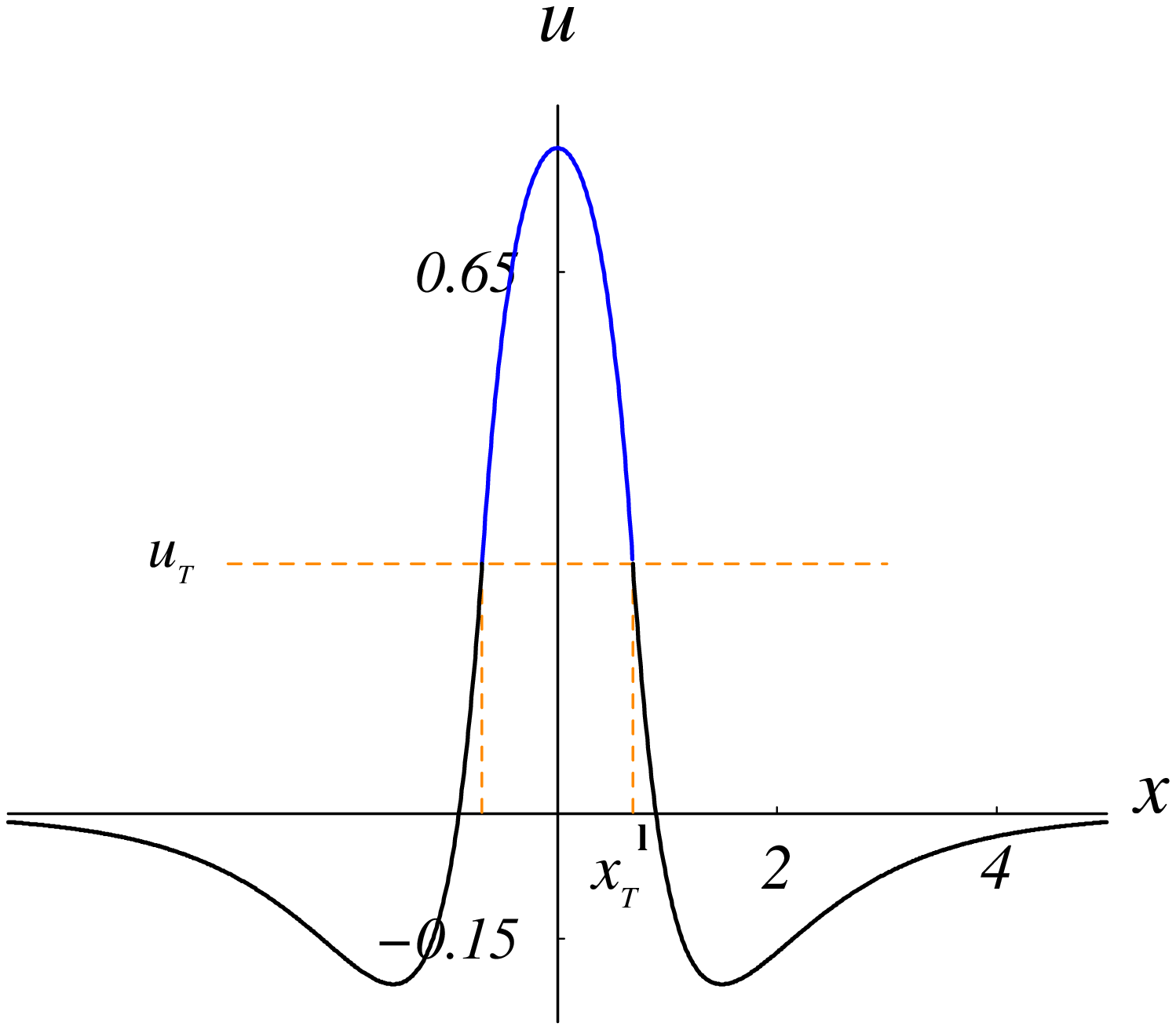, height=1.9in}
\end{minipage}
\begin{minipage}{2.5in}
\centering \epsfig{figure=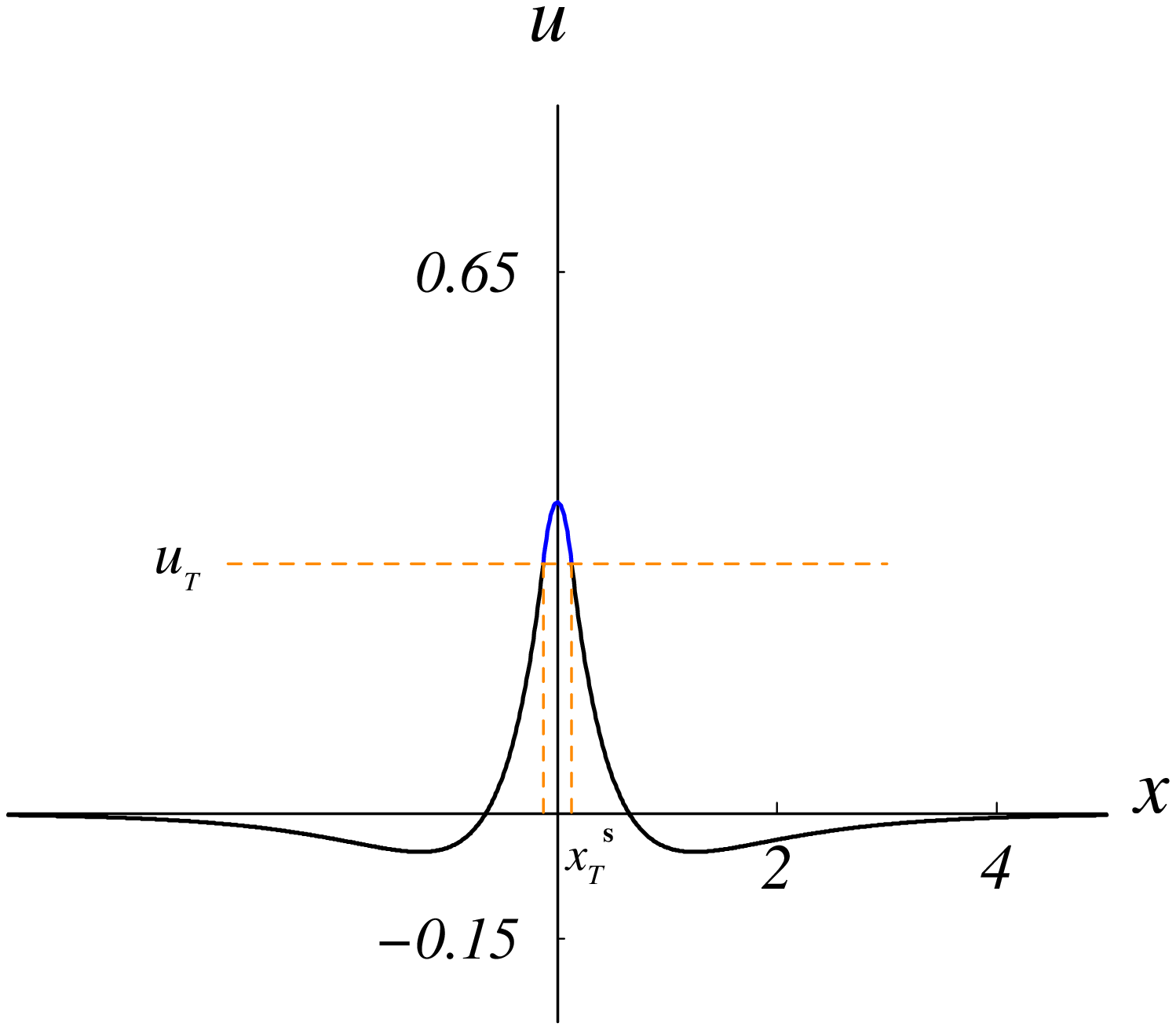, height=1.9in}
\end{minipage}
\caption{\small Large single-pulse {\bf l} and small single-pulse {\bf
s} for $A=2.8$, $a=2.6$, $\alpha=0$, $u_{\scriptscriptstyle T}=0.3$.
(Left) Single-pulse {\bf l}: $x_{\scriptscriptstyle
T}^{\scriptscriptstyle {\bf l}}=0.68633$,
height=$u(0)=0.79991$. (Right) Single-pulse {\bf s}:
$x_{\scriptscriptstyle T}^{\scriptscriptstyle {\bf s}}=0.12985$,
height=$u(0)=0.37358$.} \centering
\label{fig:ambump}
\end{figure}
The system (\ref{eq:am1}) - (\ref{eq:am5}) is linear in the
coefficients $C$, $D$, $E$, and $F$ which can be solved in terms of
$x_{\scriptscriptstyle T}$:
\[
\begin{array}{lll}
C & = & {\displaystyle -\frac{A}{a}e^{-ax_{\scriptscriptstyle T}} }\\
D & = & e^{-x_{\scriptscriptstyle T}} \\ E & = & {\displaystyle
\frac{A}{a}(e^{ax_{\scriptscriptstyle T}}-e^{-ax_{\scriptscriptstyle
T}})} \\ F & = & -(e^{x_{\scriptscriptstyle
T}}-e^{-x_{\scriptscriptstyle T}}) \\
\end{array}
\]
>From these coefficients we arrive at the following proposition for single-pulse solutions.
\begin{proposition}
\label{cl:amaripulsecondition}
There are two pulse solutions when ${u_{\scriptscriptstyle T}
\leq\int_{0}^{(\ln A)/(a-1)}w(x)dx}$ and ${\displaystyle
(\frac{A}{a}-1) < u_{\scriptscriptstyle T}}$ for $A>a$ and $0\le
u_{\scriptscriptstyle T}$ for $A<a$ .
\end{proposition}
\begin{proof}
Substituting $E$ and $F$ into (\ref{eq:am1}) (or $C$ and $D$ into
(\ref{eq:am2})), gives an existence condition for a single-pulse:
$\Phi(x_{\scriptscriptstyle T})=u_{\scriptscriptstyle T}$, where
\begin{equation}
\Phi(x)= \frac{A}{a}(1-e^{-2ax})-(1-e^{-2x}).
\end{equation}
We term $\Phi(x)$ the ``existence function''. Two examples are shown in 
Figs.~\ref{fig:amexist1} and \ref{fig:amexist2} where the curve $\Phi(x)$ crosses
$u_{\scriptscriptstyle T}$ twice, implying that there are two single-pulse
solutions.
\begin{figure}[!htb]
\centering \epsfig{figure=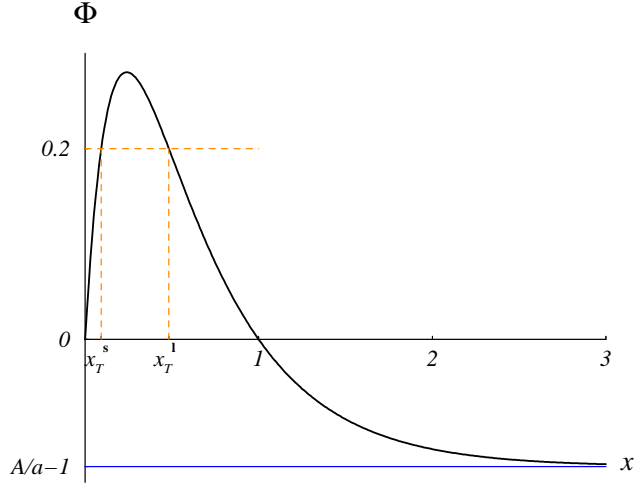, height=2.5in} 
\caption{\small
Existence function $\Phi(x)$ when $A<a$ with $\alpha=0$, $A=2.6$,
$a=3$.  $ {\displaystyle \lim_{x \rightarrow \infty}
\Phi(x)=\frac{A}{a}-1=-0.1333 }$. $\Phi(x)$ gives the range of
thresholds $ u_{\scriptscriptstyle T}$ that supports two single-pulse
solutions. Example: at $u_{\scriptscriptstyle T}=0.2$, $\Phi(x)$ shows
that we have a single-pulse solution {\bf l} with width
$x_{\scriptscriptstyle T}^{\scriptscriptstyle {\mathrm {\bf l}}}$; the
second single-pulse solution {\bf s} is narrower and has width
$x_{\scriptscriptstyle T}^{\scriptscriptstyle {\mathrm{\bf s}}}$.}
\label{fig:amexist1}
\end{figure}
\begin{figure}[!htb]
\centering \epsfig{figure=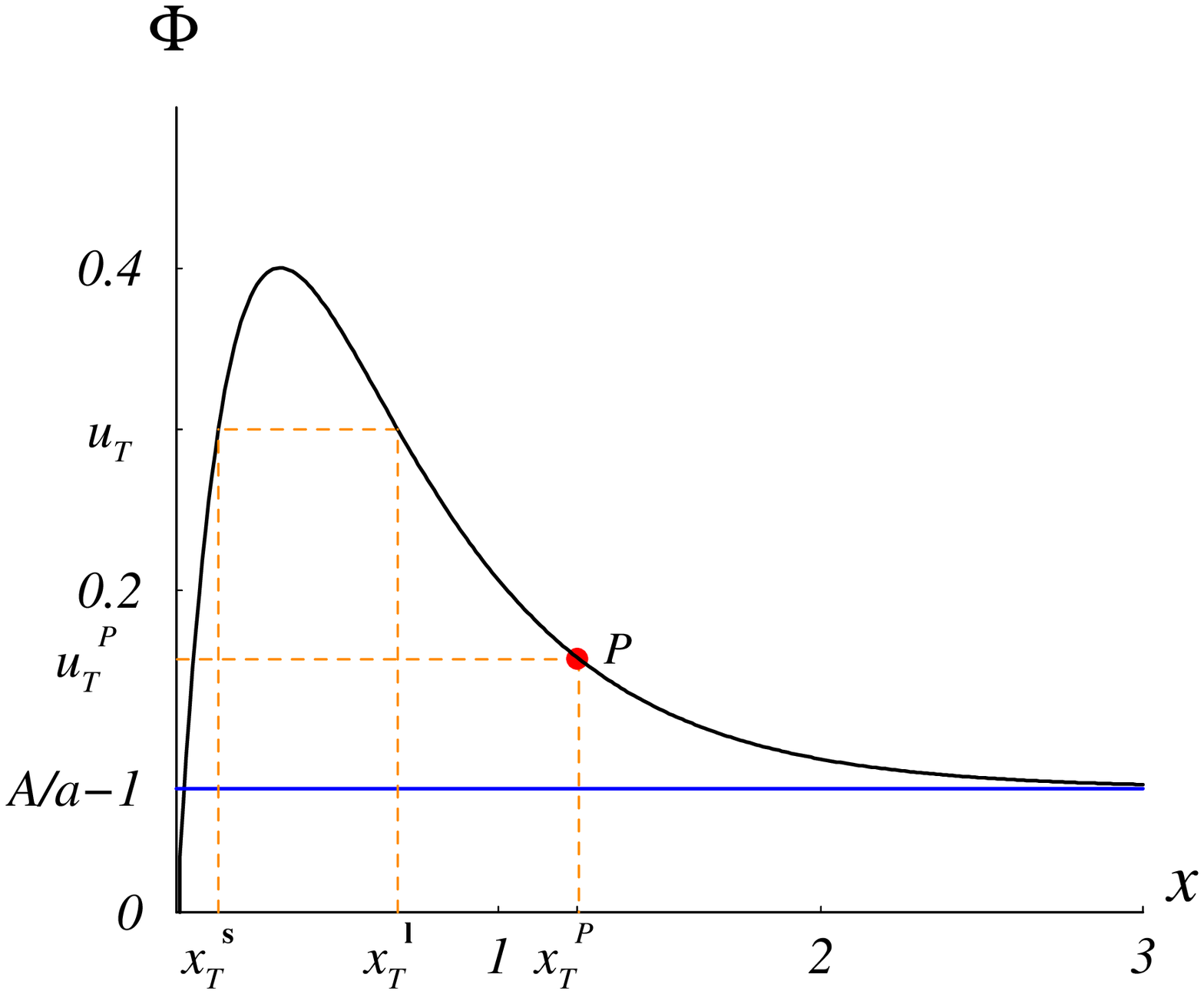, height=2.5in} \caption{\small
Existence function $\Phi(x)$ when $A>a$.  $\alpha=0$, $A=2.8$,
$a=2.6$, ${\displaystyle \lim_{x \rightarrow \infty}
\Phi(x)=\frac{A}{a}-1=0.07692.}$ Example: at $u_{\scriptscriptstyle
T}=0.3$, $\Phi(x)$ shows that there is a wide single-pulse solution {\bf l}
with width $x_{\scriptscriptstyle
T}^{\scriptscriptstyle \mathrm{\bf l}}=0.68633$ and a narrower
single-pulse solution  {\bf s} with width $x_{\scriptscriptstyle
T}^{\scriptscriptstyle \mathrm{\bf s}}=0.12985$. $P$
is the transition point where single-pulse {\bf l} changes into a
dimple-pulse {\bf d}.  At the transition, $u_{\scriptscriptstyle
T}^{\scriptscriptstyle P}$=0.15672 and  $x_{\scriptscriptstyle
T}^{\scriptscriptstyle P}=1.24379$.} 
\label{fig:amexist2}
\end{figure}
Since
$$ \lim_{x \rightarrow \infty} \Phi(x)=\frac{A}{a}-1= \left \{
\begin{array}{lcc} <0 & \mbox{if $A<a$} & (\mbox{ Figure
\ref{fig:amexist1}}) \\ \geq 0 & \mbox{if $A \geq a$} & (\mbox{Figure
\ref{fig:amexist2}})
\end{array} \right.,$$
the lower bound of $u_{\scriptscriptstyle T}$ that
supports two pulses is $0$ if $A<a$ and the lower bound of
$u_{\scriptscriptstyle T}$ that guarantees two pulses is
${\displaystyle \frac{A}{a}-1}$ when $A>a$.

The upper bound on threshold $u_{\scriptscriptstyle T}$ that supports
two pulse solutions is the maximum of $\Phi(x)$.  Solving
\begin{eqnarray*}
\frac{d \Phi}{d x}=Ae^{-2ax}-2e^{-2x}=0
\end{eqnarray*}
gives ${\displaystyle x=\frac{\ln A}{2(a-1)}}$.
Thus $\Phi$ reaches its maximum at
\begin{equation}
\Phi(x)=\frac{A}{a}(1-e^{-\frac{a \ln A}{a-1}})-(1-e^{-\frac{\ln
A}{a-1}})=\int_{0}^{\frac{\ln A}{a-1}} w(x)dx
\end{equation}
proving the proposition. \qquad
\end{proof}
 
\begin{figure}[!htb]
\centering \epsfig{figure=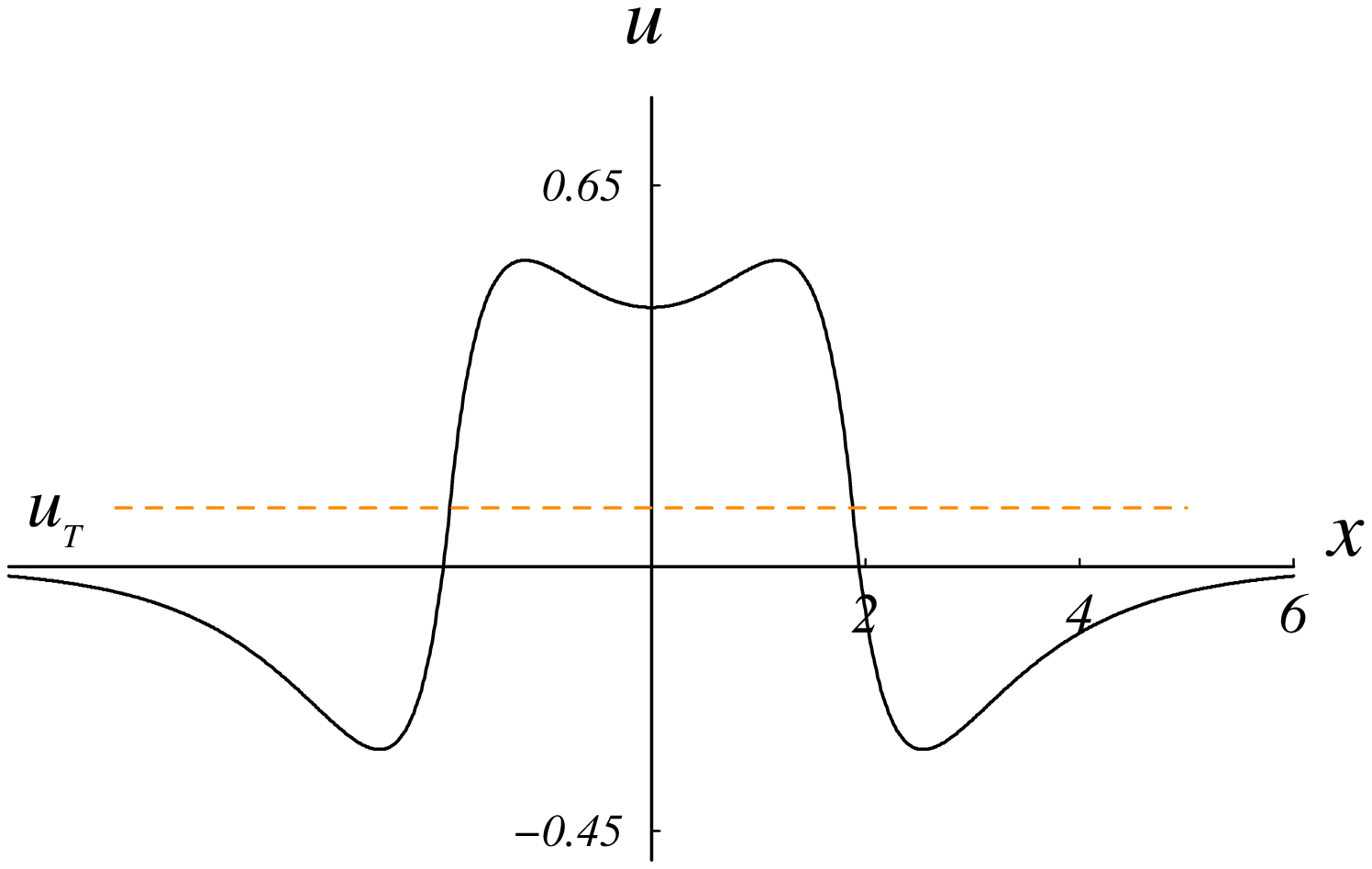, height=2.in} \caption{\small
Dimple-pulse {\bf d} for $A=2.8,$ $a=2.6,$ $\alpha=0,$ 
$u_{\scriptscriptstyle T}$ with width $x_{\scriptscriptstyle T}^{\mathrm {d}}=1.8832$.} 
\label{fig:amaridimple}
\end{figure}

Proposition \ref{cl:amaripulsecondition} does not immediately imply
that there are two single-pulses because for a
small threshold there can exist a dimple-pulse (Fig \ref{fig:amaridimple}.)
In Fig.~\ref{fig:amexist2}, as $u_T$ is lowered, $P$ is the transition
point where $u''=0$ and  single-pulse  
{\bf l} transforms into the dimple-pulse ${\bf d}$. The small
single-pulse {\bf s} always remains a 
single-pulse.  The transition point $P$ is identified by following $u''(0)$ 
as a function of $u_{\scriptscriptstyle T}$
using the continuation program AUTO. 
In the accompanying paper we compute the stability of these solutions.
In agreement with Amari~\cite{Amari77} we find that the large pulse is
stable and the small pulse is unstable.  Additionally, we find that
dimple pulse solutions can also be stable.

\subsection{Solutions for the general case}
\label{sec:eigenvaluestructure} For the general case of $\alpha \neq 0$, the
complex eigenvalues $\omega_1$, $-\omega_1$, $\omega_2$, 
$-\omega_2$ given by (\ref{eiv1}) and (\ref{eiv2}) will change form
for different parameter values. The transition
points for the eigenvalues are given by the relative signs of
functions $R$ (\ref{R}), $S$ (\ref{S}) and  $\Delta$ (\ref{Delta}).
There are three cases: both eigenvalues $\omega_{1,2}$ are
real, both are complex or both are imaginary.  

We consider the transitions when $\alpha$ is changed for fixed $a$ and
$A$.  We find that there are five critical points where the eigenvalue
structure changes.  At $\alpha=\alpha_0\equiv a/(2(A-a))$, $R-S=0$,
with $R>0$ and $\Delta>0$.  The solutions of the quadratic equation
$\Delta(\alpha)=0$ give $\alpha_1$ and $\alpha_3$.  At $\alpha_2$, $R=0.$  
At $\alpha_4=a/(2(A-a))$, $R+S=0$ with
$R<0$ and $\Delta>0$.  We arrange $\alpha_i$ ($i=0,1,2,3,4$) in
increasing order. $\omega_1$ and $\omega_2$ are complex conjugates for 
both $\alpha \in (\alpha_1, \alpha_2)$ and $\alpha \in (\alpha_2, \alpha_3).$
In our analysis, we only consider the case where 
$\alpha>0$ (i.e. the firing rate is increasing with input).  The case of
$\alpha=0$ with the general connection weight function was fully treated
in \cite{Amari77} and is reevaluated in Sec.~\ref{sec:amari}.
Tables~\ref{tab:eigenvalue1}, \ref{tab:eigenvalue2}, and
\ref{tab:eigenvalue3} enumerate all the possible forms of $\omega_1$
and $\omega_2$.
\begin{table}[hbt!]
\centering
\caption{Eigenvalue chart when $A>a$}
\begin{small}
\begin{tabular}{|c|c|c|c|c|c|c|} \hline
   & E1 & E2 & E3 & E4& E5& E6 \\ \hline & $\triangle >0$ & $\triangle
   >0$ & $\triangle >0$ & $\triangle=0$ & $\triangle<0$ & $\Delta>0$\\
   &$R<0<|R|$ & &$R<0<S$ & & & $R<0<|R|$ \\ & $|R|<S$ & $0<S<R$
   &$S<|R|$ & & & $|R|=S$ \\ \hline $ \omega_1$ & real & real$$
   & imaginary$$ & $=\omega_2$ &$=\omega_2^*$, complex &
   $=\sqrt{2R}$\\ \hline $\omega_2$& imaginary & real $$ & imaginary$$
   & $=\omega_1$ &$=\omega_1^*$, complex &$0$ \\ \hline $\alpha$ &
   $(\alpha_{\scriptscriptstyle 4},\infty)$ &
   $(-\infty,\alpha_{\scriptscriptstyle 1})$ &
   $(\alpha_{\scriptscriptstyle 3},\alpha_{\scriptscriptstyle 4})$ &
   $\alpha_{\scriptscriptstyle 1},\alpha_{\scriptscriptstyle 3}$ &
   $(\alpha_{\scriptscriptstyle 1},\alpha_{\scriptscriptstyle 3})$ &
   $\alpha_{\scriptscriptstyle 4}$ \\ \hline
\end{tabular}
\end{small}
\label{tab:eigenvalue1}
\end{table}
\begin{table}[hbt!]
\centering
\caption{Eigenvalue chart when $A<a$.}
\begin{small}
\begin{tabular}{|c|c|c|c|c|c|c|} \hline
   & E1 & E2 & E3 & E4& E5& E6 \\ \hline & $\triangle >0$ & $\triangle
   >0$ & $\triangle >0$ & $\triangle=0$ & $\triangle<0$ & $\Delta>0$\\
   & $0<R<S$ & $0<S<R$ &$R<0<S<|R|$ & & & $0<R=S$ \\ \hline $
   \omega_1$ & real & real$$ & imaginary$$ & $=\omega_2$
   & complex & $=\sqrt{2R}$\\ \hline $\omega_2$&
   imaginary & real $$ & imaginary$$ & $=\omega_1$ &,
   complex &$0$ \\ \hline $\alpha$ &
   $(-\infty,\alpha_{\scriptscriptstyle 0})$ &
   $(\alpha_{\scriptscriptstyle 0},\alpha_{\scriptscriptstyle 1})$ &
   $(\alpha_{\scriptscriptstyle 3},\infty)$ &
   $\alpha_{\scriptscriptstyle 1},\alpha_{\scriptscriptstyle 3}$ &
   $(\alpha_{\scriptscriptstyle 1},\alpha_{\scriptscriptstyle 3})$ &
   $\alpha_{\scriptscriptstyle 0}$ \\ \hline
\end{tabular}
\end{small}
\label{tab:eigenvalue2}
\end{table}
\begin{table}[hbt!]
\caption{Eigenvalue chart when $A=a$.}
\centering
\begin{small}
\begin{tabular}{|c|c|c|c|c|c|c|} \hline
   & E1 & E2 & E3 & E4& E5& E6 \\ \hline & $\triangle >0$ & $\triangle
   >0$ & $\triangle >0$ & $\triangle=0$ & $\triangle<0$ & \\
 & $0<R<S$ & $0<S<R$ &$R<0<S<|R|$ & & & $R=S$ \\ \hline $
\omega_1$ & & real$$ &imaginary & $=\omega_2$ &$=\omega_2^*$, complex&
$$\\ \cline{1-1} \cline{3-6} \cline{5-6} $\omega_1$& $\emptyset$& real
$$ &$$imaginary & $=\omega_2$ &$=\omega_1^*$, complex &$ \emptyset$ \\
\cline{1-1} \cline{3-6} $\alpha$ & $$ &
$(-\infty,\alpha_{\scriptscriptstyle 1})$ &
$(\alpha_{\scriptscriptstyle 3},\infty)$ & $\alpha_{\scriptscriptstyle
1},\alpha_{\scriptscriptstyle 3}$ & $(\alpha_{\scriptscriptstyle
1},\alpha_{\scriptscriptstyle 3})$ & $$\\ \hline
\end{tabular}
\end{small}
\label{tab:eigenvalue3}
\end{table}

Although both $\alpha_0$ and $\alpha_4$ have the same expression, they
do not co-exist.  When $A<a$, ${\displaystyle
\alpha_0=\frac{a}{2(A-a)}<0}$ and when $A>a$, ${\displaystyle
\alpha_4=\frac{a}{2(A-a)}>0}$.
For all values of $\omega_1$ and $\omega_2$,
  $u_{\mathrm {II}}(x)$ and $u_{\mathrm {III}}(x)$ always have the
  form (\ref{eqa:solregII}) and (\ref{eqa:solregIII}) respectively.

\newpage
\subsection{Solutions for real eigenvalues}

\subsubsection{Construction of single-pulse solutions}
\label{sec:real}

\begin{figure}[!htb]
\centering \epsfig{figure=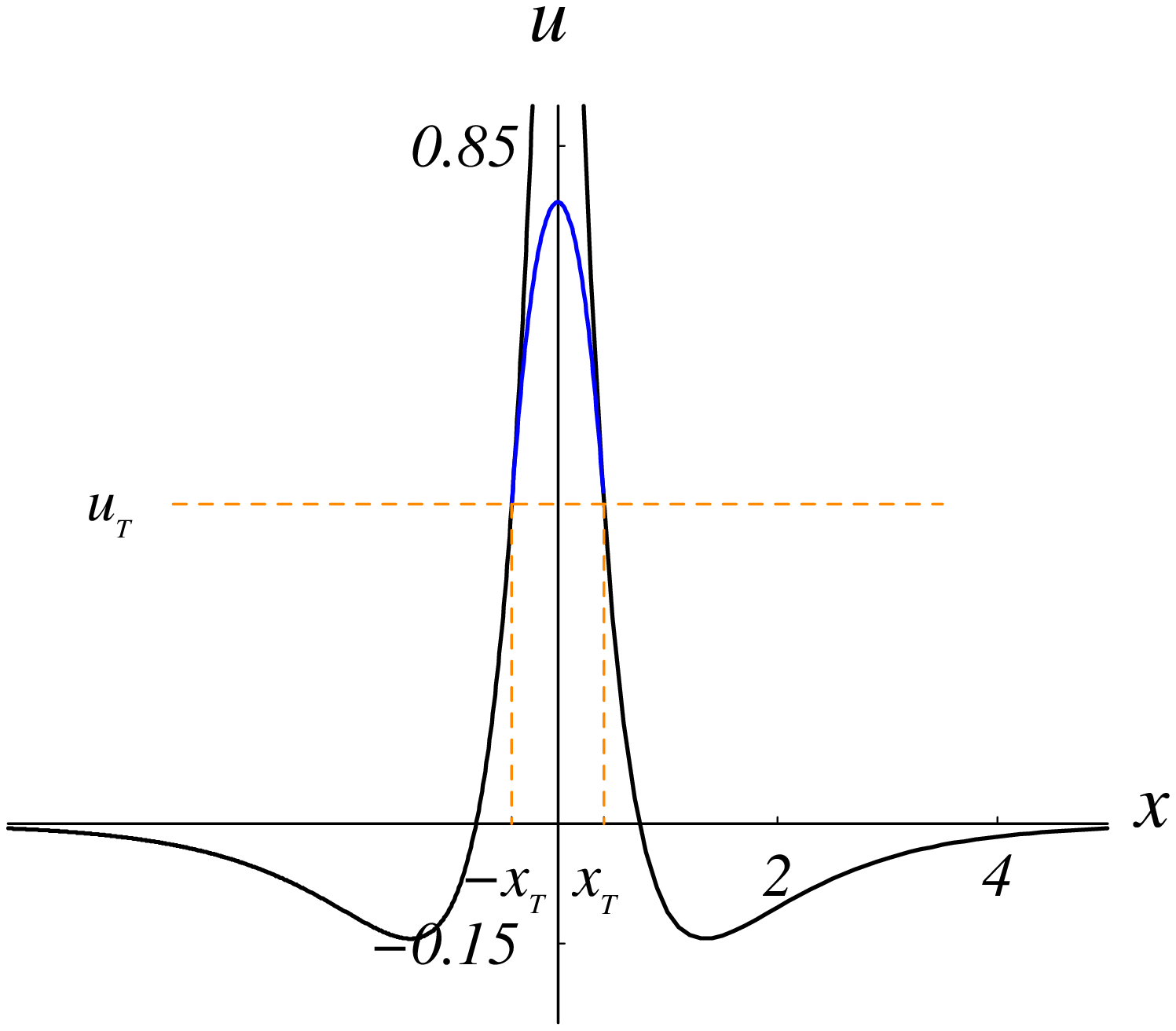, height=2.5in}
\caption{\small Construction of large single-pulse {\bf l}. $A=2.8$,
  $a=2.6$, $\alpha=0.15$, $u_{\scriptscriptstyle T}=0.3.$
  $x_{\scriptscriptstyle 
T}^{\scriptscriptstyle \mathrm {\bf l}}=0.41092$, height=$u(0)=0.77892$.} 
\centering
\label{fig:0.15alpbumpmatch}
\end{figure}

\begin{figure}[!htb]
\begin{minipage}{2.5in}
\centering \epsfig{figure=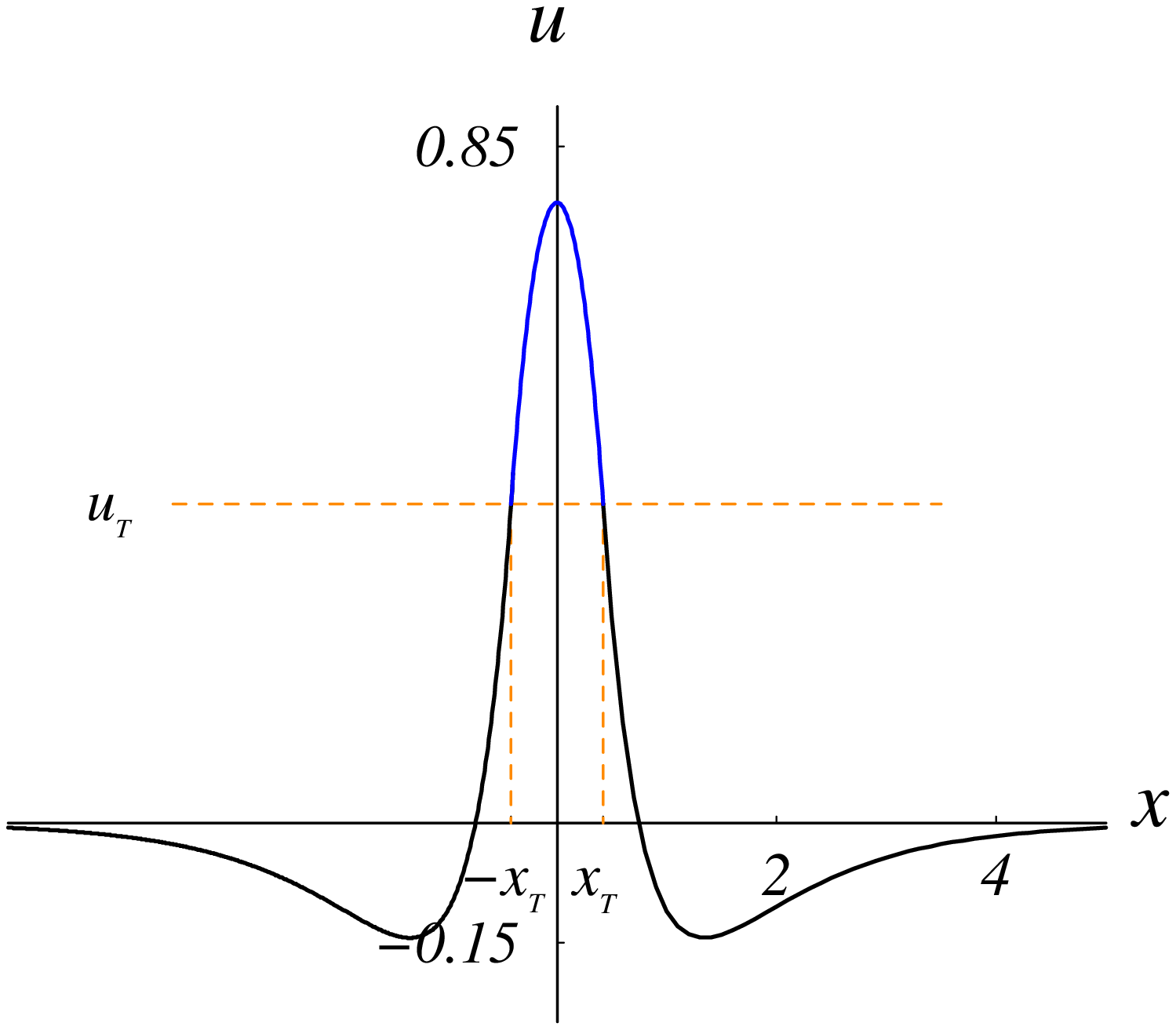, height=2.in}
\end{minipage}
\begin{minipage}{2.5in}
\centering \epsfig{figure=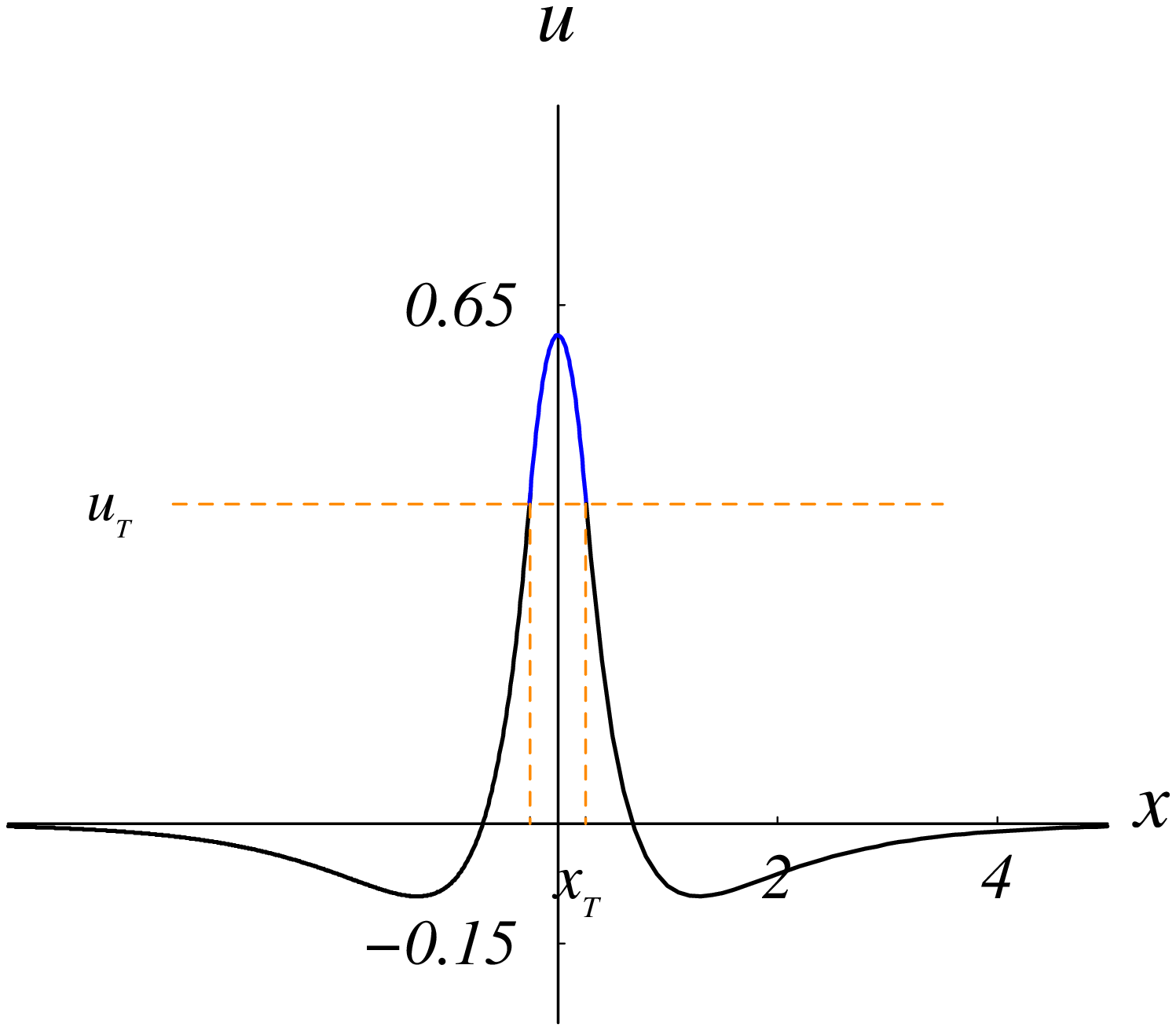, height=2.in}
\end{minipage}
\caption{\small Large single-pulse {\bf l} and small single-pulse {\bf
s}. $A=2.8$, $a=2.6$, $\alpha=0.15$, $u_{\scriptscriptstyle
T}=0.3$. (Left) Single-pulse {\bf l}: $x_{\scriptscriptstyle
T}^{\scriptscriptstyle \mathrm {\bf l}}=0.41092$,
height=$u(0)=0.77892$. (Right) Single-pulse {\bf s}:
$x_{\scriptscriptstyle T}^{\scriptscriptstyle \mathrm{\bf s}}=0.2582$,
height=$u(0)=0.6123$.}
\label{fig:0.15alpsmallbigbump} \centering
\end{figure}

For $\alpha \in (0,\alpha_1)$, both $\Delta$ and $R$ are 
positive, so $\omega_1$ and $\omega_2$ are both real.  Hence
$u_{\mathrm I}(x)$ and $u_{\mathrm {II}}(x)$ have the following form:
\begin{eqnarray}
\label{eq:solregI}
u_{\mathrm I}(x) & = &
C(e^{\omega_1x}+e^{-\omega_1x})+D(e^{\omega_2x}+e^{-\omega_2x})+\frac{2(A-a)(\beta-\alpha
u_{\scriptscriptstyle T})}{a-2\alpha(A-a)}\\
\label{eq:solregII}
u_{\mathrm {II}}(x)& = & Ee^{-ax}+Fe^{-x}
\end{eqnarray}
When eigenvalues $\omega_1$ and $\omega_2$ are real, $C$ and $D$ must
also be real for real $u_{\mathrm I}(x)$. Applying the the matching conditions
(\ref{eq:matchingcondition1})-(\ref{eq:matchingcondition5}) to 
(\ref{eq:solregI}) and (\ref{eq:solregII}) yields
\begin{small}
\begin{eqnarray}
\label{eq:real1}
Ee^{-ax_{\scriptscriptstyle T}}+Fe^{-x_{\scriptscriptstyle T}}& =&
u_{\scriptscriptstyle T} \\
\label{eq:real2}  C(e^{\omega_1x_{\scriptscriptstyle
T}}+e^{-\omega_1x_{\scriptscriptstyle
T}})+D(e^{\omega_2x_{\scriptscriptstyle
T}}+e^{-\omega_2x_{\scriptscriptstyle T}})+U_0 &=& u_{\scriptscriptstyle
T} \\ \label{eq:real3} \omega_1C(e^{\omega_1x_{\scriptscriptstyle
T}}-e^{-\omega_1x_{\scriptscriptstyle
T}})+\omega_2D(e^{\omega_2x_{\scriptscriptstyle
T}}-e^{-\omega_2x_{\scriptscriptstyle T}}) &=&
-aEe^{-ax_{\scriptscriptstyle T}}-Fe^{-x_{\scriptscriptstyle T}} \\
\label{eq:real4} \omega_1^2C(e^{\omega_1x_{\scriptscriptstyle
T}}+e^{-\omega_1x_{\scriptscriptstyle
T}})+\omega_2^2D(e^{\omega_2x_{\scriptscriptstyle
T}}+e^{-\omega_2x_{\scriptscriptstyle T}})& =&
a^2Ee^{-ax_{\scriptscriptstyle T}}+Fe^{-x_{\scriptscriptstyle T}}- \\ 
& & 2 (a A-1)\beta \nonumber\\
\label{eq:real5}
\hspace{1.3cm}\omega_1^3C(e^{\omega_1x_{\scriptscriptstyle
T}}-e^{-\omega_1x_{\scriptscriptstyle
T}})+\omega_2^3D(e^{\omega_2x_{\scriptscriptstyle
T}}-e^{-\omega_2x_{\scriptscriptstyle T}})& =& (-a^3+2a \alpha
(aA-1))Ee^{-ax_{\scriptscriptstyle T}}+ \\
 & &(-1 +2 \alpha (aA-1))Fe^{-x_{\scriptscriptstyle T}}\nonumber
\end{eqnarray}
\end{small}
System
(\ref{eq:real1}) - (\ref{eq:real5}) can be solved for the five
unknowns $C, D, E, F$  and $x_{\scriptscriptstyle T}$ using Mathematica
\cite{Mathematica}, giving an explicit formula for
$u_{\mathrm I}(x)$ and $u_{\mathrm {II}}(x)$.  The single-pulse is
then given by $u_{\mathrm I}(x)$ on
$(-x_{\scriptscriptstyle T}, x_{\scriptscriptstyle T})$, $u_{\mathrm
{II}}(x)$ on $(x_{\scriptscriptstyle T}, \infty)$ and $u_{\mathrm
{III}}(x)$ on $(-\infty, x_{\scriptscriptstyle T})$. 
For the parameter set
$(a, A, \alpha, \beta, u_{\scriptscriptstyle T})=(2.6, 2.8, 0.15,
1, 0.400273)$, the solution is $(C, D, E, F, x_{\scriptscriptstyle T}
)=(-0.8532, 1.16865, 2.94108, -0.89571, 0.41902)$. 
Figure~\ref{fig:0.15alpbumpmatch} shows a graph of this single-pulse.
The height
$u_{\mathrm I}(0)$ of the pulse is $0.77892$. Its width is
$x_{\scriptscriptstyle T}^{\scriptscriptstyle \mathrm{\bf
l}}=0.41902$.  There also exists a second smaller and narrower
single-pulse solution  to (\ref{eq:real1})-(\ref{eq:real5}) for the same 
set of parameters (see Fig.~\ref{fig:0.15alpsmallbigbump}). The  
height and the width of this pulse are $u_{\mathrm I}(0)=0.6123$ and
$x_{\scriptscriptstyle T}^{\scriptscriptstyle \mathrm{\bf s}}=0.2582$
respectively.


When $\Delta=0$ and $R>0$ ($\alpha=\alpha_1$), there are repeating
real eigenvalues: $\omega_1$, $\omega_1$, $-\omega_1$, $-\omega_1$,
where $\omega_1=\sqrt{R}$.  By the symmetry of $u_{\mathrm
I}(x)$, $C1=D1$ and $C2=D2$, so  $u_{\mathrm I}(x)$ from
(\ref{eq:generalsolregI}) 
can be written as
\begin{eqnarray}
u_{\mathrm
I}(x)=C(e^{\omega_1x}+e^{-\omega_1x})
+ Dx(e^{\omega_1x}+e^{-\omega_1x})+U_0.
\end{eqnarray}
Applying matching conditions
($\ref{eq:matchingcondition1}$)-($\ref{eq:matchingcondition5}$) gives
a similar system to (\ref{eq:real1})-(\ref{eq:real5}) to which
solutions can be found numerically.

\subsubsection{Finding solutions for real eigenvalues using the
  existence function} 
\label{sec:realexistencefunction}

In order to perform numerical continuation on the single-pulse
solutions,  it is more convenient to utilize the existence function $\Phi(x)$
introduced by  Amari~\cite{Amari77} and calculated in Sec.~\ref{sec:amari}. 
We compute it for the general case by first
eliminating the threshold $u_{\scriptscriptstyle T}$ from system
(\ref{eq:real1})-(\ref{eq:real5}) to get an equivalent four equation
system
\begin{small}
\begin{eqnarray}
\label{eq:ex1}\hspace{1cm} C(e^{\omega_1x_{\scriptscriptstyle
T}}+e^{-\omega_1x_{\scriptscriptstyle
T}})+D(e^{\omega_2x_{\scriptscriptstyle
T}}+e^{-\omega_2x_{\scriptscriptstyle
T}})+U_0 &=& \\
&&\hspace{-1.6cm} \frac{a}{a-2\alpha(A-a)}
(Ee^{-ax_{\scriptscriptstyle T}}+Fe^{-x_{\scriptscriptstyle T}}) \nonumber\\
\label{eq:ex2} 
\omega_1C(e^{\omega_1x_{\scriptscriptstyle
T}}-e^{-\omega_1x_{\scriptscriptstyle
T}})+\omega_2D(e^{\omega_2x_{\scriptscriptstyle
T}}-e^{-\omega_2x_{\scriptscriptstyle T}}) & =&
-aEe^{-ax_{\scriptscriptstyle T}}-Fe^{-x_{\scriptscriptstyle T}} \\
\label{eq:ex3} 
\omega_1^2C(e^{\omega_1x_{\scriptscriptstyle
T}}+e^{-\omega_1x_{\scriptscriptstyle
T}})+\omega_2^2D(e^{\omega_2x_{\scriptscriptstyle
T}}+e^{-\omega_2x_{\scriptscriptstyle T}}) &=&
a^2Ee^{-ax_{\scriptscriptstyle T}}+Fe^{-x_{\scriptscriptstyle T}}- \\ 
& & 2(aA-1)\beta \nonumber\\
\omega_1^3C(e^{\omega_1x_{\scriptscriptstyle
T}}-e^{-\omega_1x_{\scriptscriptstyle
T}})+\omega_2^3D(e^{\omega_2x_{\scriptscriptstyle
T}}-e^{-\omega_2x_{\scriptscriptstyle T}}) &=& (-a^3+2a \alpha
(aA-1))Ee^{-ax_{\scriptscriptstyle T}}+ \\ & & (-1 +2 \alpha
(aA-1))Fe^{-x_{\scriptscriptstyle T}} \nonumber
\label{eq:ex4}
\end{eqnarray}
\end{small}
Equations (\ref{eq:ex1})-(\ref{eq:ex4}) form a linear system in $C$,
$D$, $E$, and $F$. To obtain an existence function $\Phi(x)$, we construct
coefficient vectors
\bigskip

 $m_1=\left( \begin{array}{c} e^{\omega_1x_{\scriptscriptstyle
 T}}+e^{-\omega_1x_{\scriptscriptstyle T}}\\
 \omega_1(e^{\omega_1x_{\scriptscriptstyle
 T}}-e^{-\omega_1x_{\scriptscriptstyle T}}) \\
 \omega_1^2(e^{\omega_1x_{\scriptscriptstyle
 T}}+e^{-\omega_1x_{\scriptscriptstyle T}}) \\
 \omega_1^3(e^{\omega_1x_{\scriptscriptstyle
 T}}-e^{-\omega_1x_{\scriptscriptstyle T}})
\end{array} \right),$
$m_2=\left( \begin{array}{c} e^{\omega_2x_{\scriptscriptstyle
T}}+e^{-\omega_2x_{\scriptscriptstyle T}} \\
\omega_2(e^{\omega_2x_{\scriptscriptstyle
T}}-e^{-\omega_2x_{\scriptscriptstyle T}}) \\
\omega_2^2(e^{\omega_2x_{\scriptscriptstyle
T}}+e^{-\omega_2x_{\scriptscriptstyle T}}) \\
\omega_2^3(e^{\omega_2x_{\scriptscriptstyle
T}}-e^{-\omega_2x_{\scriptscriptstyle T}})
\end{array} \right),$
\bigskip

$m_3=\left( \begin{array}{c} \frac{a}{a -2\alpha( A-a )}\\ a\\ -a^2\\
a^3-2a\alpha (aA-1) \end{array} \right),$ $m_4=\left( \begin{array}{c}
\frac{a}{a -2\alpha(A-a)}\\ 1\\ -1\\ 1-2 \alpha (aA-1) \end{array}
\right),$ 
\bigskip

$m_0=\left( \begin{array}{c} \frac{(A-a)
\beta}{a-2\alpha(A-a)}\\ 0\\ -2(aA-1) \beta \\ 0 \end{array} \right).$
\bigskip

\begin{figure}[!htb]
\centering \epsfig{figure=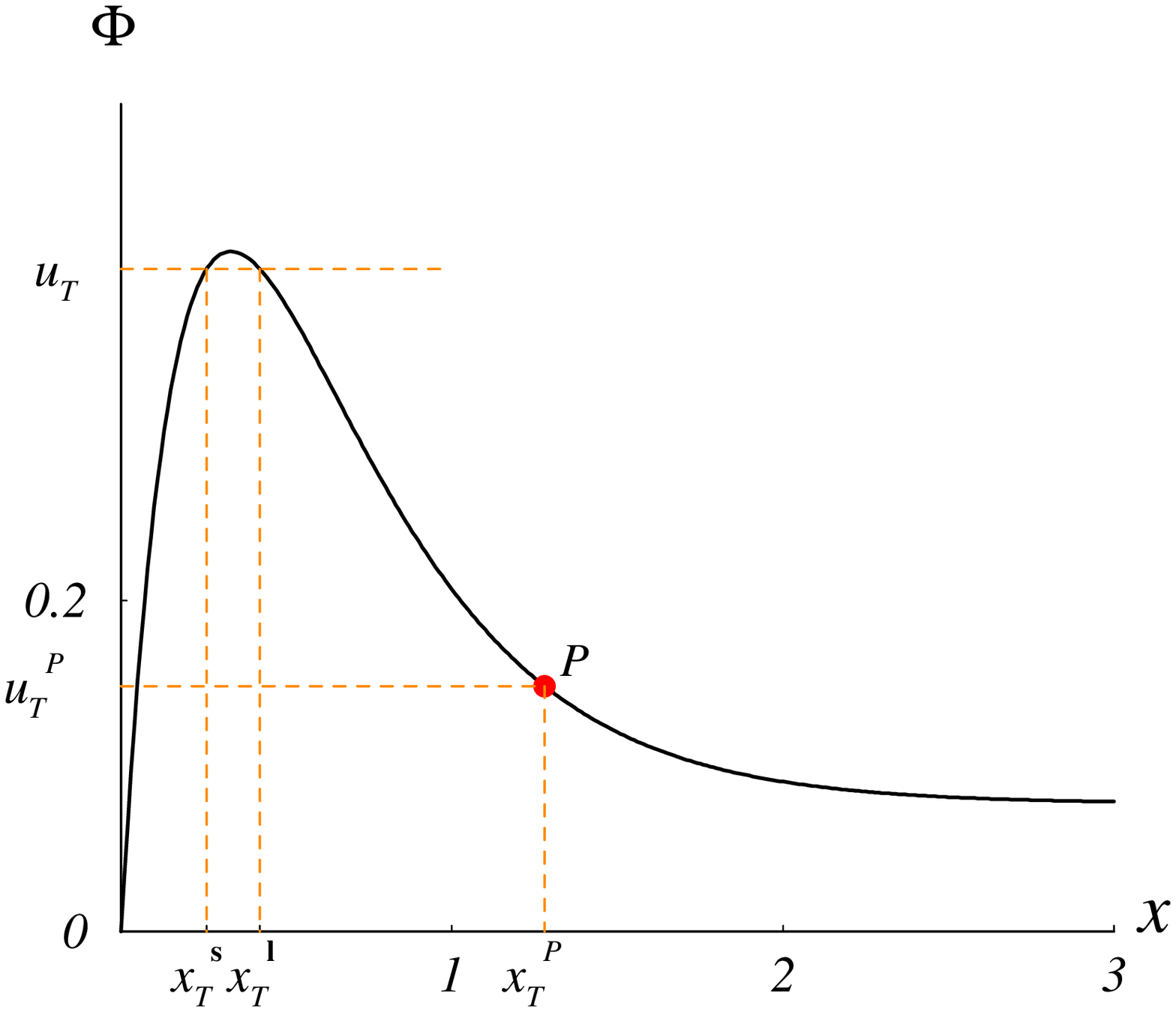, height=2.5in}
\caption{\small Existence function $\Phi(x)$ for $\alpha=0.15$, $A=2.8$,
$a=2.6$. At $u_{\scriptscriptstyle T}=0.400273$, $\Phi(x)$ has a
single-pulse {\bf l} which has width $x_{\scriptscriptstyle
T}^{\scriptscriptstyle {\bf l}}=0.4109$; the second single-pulse {\bf
s} is narrower with width $x_{\scriptscriptstyle
T}^{\scriptscriptstyle {\bf s}}=0.2582$. At $P$, threshold
$u_{\scriptscriptstyle T}^{\scriptscriptstyle P}=0.1489$, $u''(0)$ of
the pulse at $P$ is $0$.} \centering
\label{fig:0.15alpexistence}
\end{figure}

\noindent Let $DET_{ x_{\scriptscriptstyle T}}(\alpha)=\left|
\begin{array}{cccccc} & m_1 & m_2 & m_3 & m_4 & \end{array}\right|,$
where $\left|.\right|$ is the determinant.  For
parameters $(a, A, \alpha, \beta, u_{\scriptscriptstyle
T})=(2.6,2.8,0.15,1,0.400273)$, the solution $(C, D, E, F,
x_{\scriptscriptstyle T})=(-0.8532, 1.16865,$ $ 2.94108, -0.89571,
0.41902)$ with $DET_{ x_{\scriptscriptstyle
T}}(\alpha)=-243.2415568475$  is given by Mathematica. We
use this solution as an 
initial guess to continue system (\ref{eq:real1})-(\ref{eq:real5})
using AUTO while following $DET_{ x_{\scriptscriptstyle T}}(\alpha)$
as $\alpha$ decreases to $0$ and then increases to $\alpha_1$. The
recorded value of $DET_{ x_{\scriptscriptstyle T}}$ shows that
$DET_{x_{\scriptscriptstyle T}}(\alpha) \neq 0$ as
$\alpha<\alpha_1$. Therefore we can always solve the linear system
(\ref{eq:ex1})-(\ref{eq:ex4}) by Cramer's rule.  The solutions for $E$
and $F$ given by
$$E= \frac{\left|\begin{array}{cccccc} & m_1 & m_2 & m_0 & m_4 &
\end{array} \right|} {\left |\begin{array}{cccccc} & m_1 & m_2 & m_3 &
m_4 & \end{array} \right| e^{-ax_{\scriptscriptstyle T}}} ,$$
$$F= \frac{\left|\begin{array}{cccccc} & m_1 & m_2 & m_3 & m_0 &
\end{array} \right|} {\left| \begin{array}{cccccc} & m_1 & m_2 & m_3 &
m_4 & \end{array} \right| e^{-x_{\scriptscriptstyle T}} }.$$
are then substituted back into $u_{\scriptscriptstyle
II}(x)=E(x)e^{-ax}+F(x)e^{-x}$ to obtain the existence function
\begin{equation}
\Phi(x)=E e^{-ax}+F e^{-x}=\frac{\left|\begin{array}{cccccc} & m_1 &
m_2 & m_0 & (m_3-m_4) &
\end{array} \right|}{\left|\begin{array}{cccccc} & m_1 & m_2 & m_3 & m_4
& \end{array}\right|}.
\label{existfnc}
\end{equation}
Figure~\ref{fig:0.15alpexistence} shows $\Phi(x)$ (\ref{existfnc}).
$\Phi(x)$ approaches a limit as $x \rightarrow \infty,$ but this limit is 
different from the limit of $\Phi(x)$ in the Amari case ($\alpha=0$.)

Single-pulses of width $x_T$  are given by solutions of $\Phi(x_T)=u_T$.
Since all four eigenvalues are real, there are no oscillations in
$\Phi(x)$ and so there are at most two pulse
solutions. One is a small single-pulse, and the other is either a
large single-pulse or a dimple-pulse depending on the threshold
$u_{\scriptscriptstyle T}.$

\subsubsection{Transition point $P$ between single-pulses and dimple-pulses}
\label{sec:transition}
\begin{figure}[!htb]
\centering \epsfig{figure=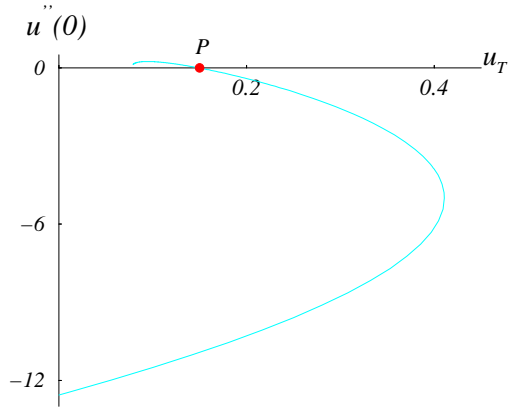, height=2in} \caption{\small
Plot of $u''(0)$ when $\alpha=0.15$, $A=2.8$, $a=2.6$. $P$ is the
transition point between single-pulse ${\bf l}$ and dimple-pulse ${\bf
d}$. When threshold $u_{\scriptscriptstyle T}^{\scriptscriptstyle
P}=0.1489$, $u''(0)=0$.} \centering
\label{fig:u2utreal}
\end{figure}

\begin{figure}[!htb]
\centering \epsfig{figure=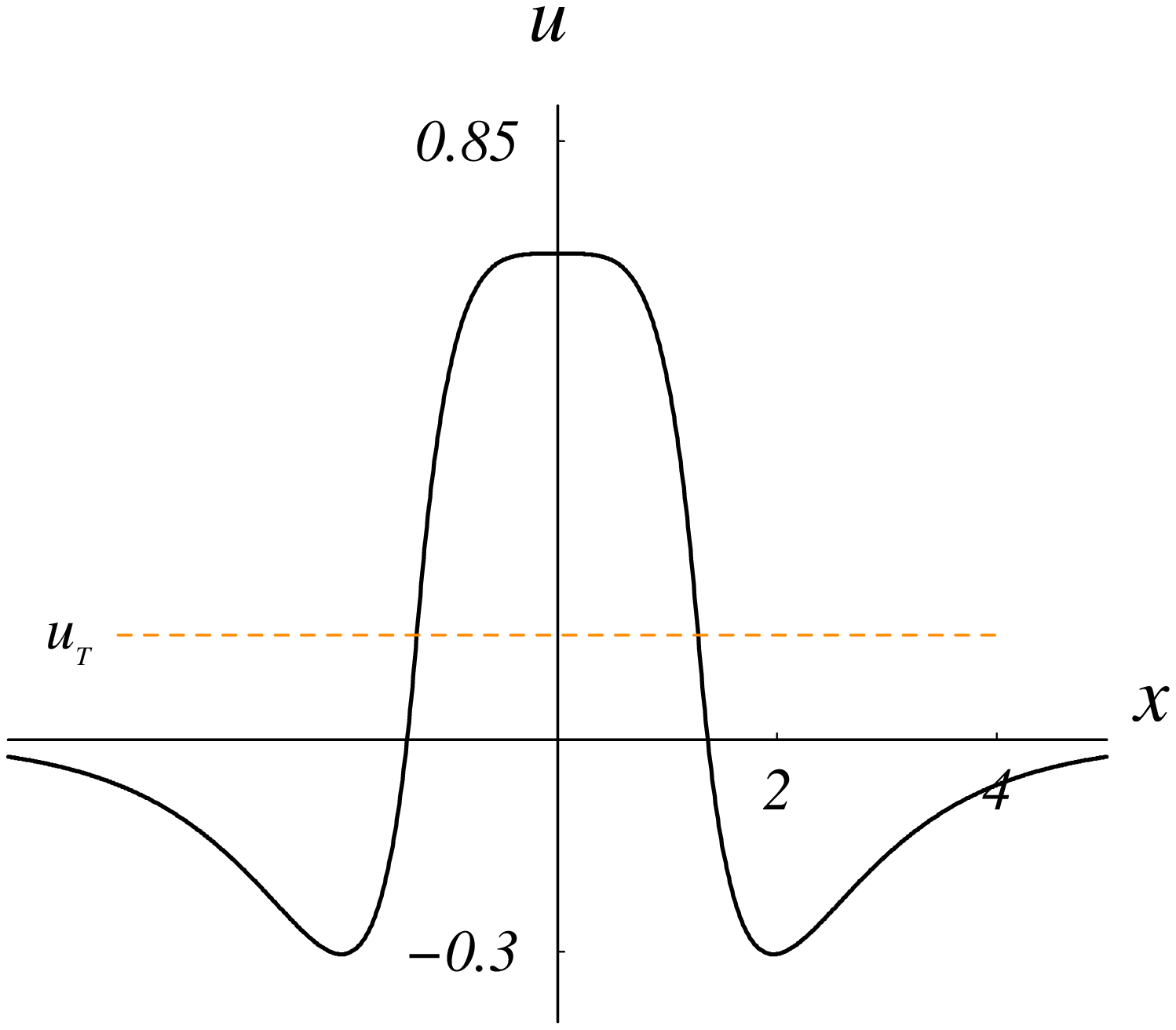, height=2in}
\caption{\small Example of P-pulse with $a=2.6$, $A=2.8$, $\alpha=0.15$,
$u_{\scriptscriptstyle T}^{\scriptscriptstyle P}=0.14838$.  The width
of this pulse is $x_{\scriptscriptstyle T}^{\scriptscriptstyle
P}=1.27978$ and $u''(0)=0$.} \centering
\label{fig:realPbump}
\end{figure}

The existence function $\Phi(x)$ gives a range of thresholds
$u_{\scriptscriptstyle T}$ for which there exist two pulse solutions;
a large pulse {\bf l} (or dimple-pulse {\bf d}) and a small pulse {\bf
s}, or only one small single-pulse solution. The $x$-value of the
intersection of $u_{\scriptscriptstyle T}$ and $\Phi(x)$ is the width
of a pulse. In figure \ref{fig:0.15alpexistence},
$x_{\scriptscriptstyle T}^{\scriptscriptstyle {\bf s}}$ is the width
of {\bf s}, and $x_{\scriptscriptstyle T}^{\scriptscriptstyle {\bf
l}}$ is the width of {\bf l}. At $P$,  the curvature at the peak
of the pulse solution is zero (i.e. $u''(0)=0$)
as seen in Fig.~\ref{fig:realPbump}. For this
set of parameters, dimple-pulses
appear if $u_T$ is between $u_{\scriptscriptstyle
T}^{P}=0.1489$ and  $\lim_{x\rightarrow\infty}\Phi(x)$
(See Fig.~\ref{fig:0.15alpexistence}). Figure \ref{fig:u2utreal} shows the
continuation plot of $u''(0)$; $u''(0)$ crosses
zero at $u_{\scriptscriptstyle T}=u_{\scriptscriptstyle T}^P.$

\subsubsection{Loss of existence for unbalanced synaptic connectivity} 
\label{sec:toomuchexcitation}

\begin{figure}[htb!]
\centering \epsfig{figure=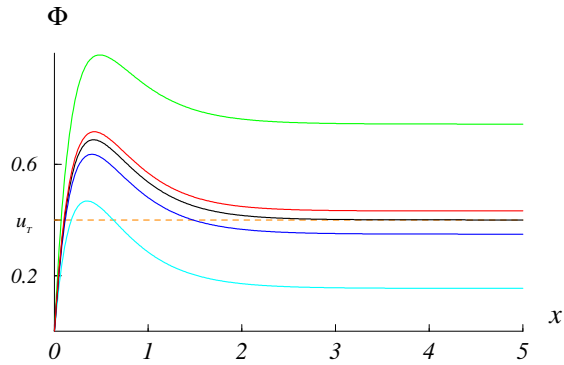, height=1.8in}
\caption{\small $\Phi(x)$ with excitation dominating
  inhibition. $a=2.6,$ $\alpha=0.05,$ $u_{\scriptscriptstyle T} 
=0.400273,$ and different values of $A$: $A=3$ (cyan); $A=3.5$ (blue); $A=3.62$
(black); $A=3.7$ (red); $A=4.5$ (green). When $\Phi(x)$ is tangential to 
threshold $u_{\scriptscriptstyle T}$ for large $x,$ the width of the large 
pulse is $\infty$.}
\centering
\label{fig:bigexcitation}
\end{figure}

\begin{figure}[htb!]
\centering \epsfig{figure=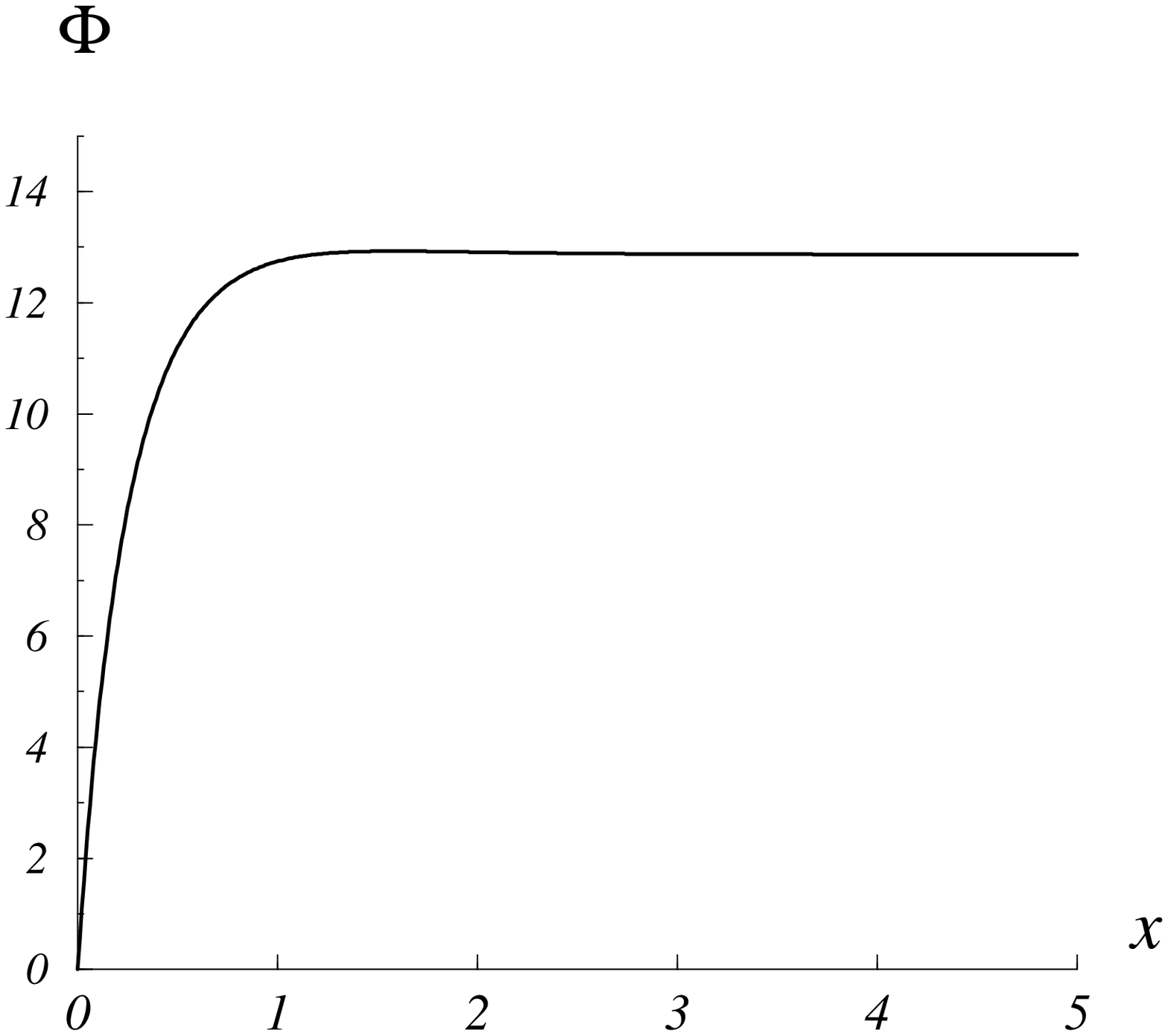, height=1.6in}
\caption{\small $\Phi(x)$ with excitation dominating inhibition: $A=29.6$,
  $a=2.6$, $\alpha=0.03$.  There exists only pulse {\bf s}, pulse {\bf l} or
	{\bf d} do not exist.}
\centering
\label{fig:newAaexist}
\end{figure}

An examination of the shape of $\Phi(x)$ shows why single-pulses do
not exist when excitation and inhibition are too much out of balance.
When excitation dominates, $A/a>1$. For fixed $a,$ $\alpha$ and threshold 
$u_{\scriptscriptstyle T},$ as $A$ becomes larger,
the existence function $\Phi(x)$ moves up and 
${\displaystyle \lim_{x \rightarrow
 \infty}\Phi(x)}$ becomes larger. The width of the large pulse {\bf l} (or dimple-pulse) 
increases (Fig \ref{fig:bigexcitation}.) When $\Phi(x)$ is tangent to 
$u_{\scriptscriptstyle T}$ for large $x$ (black curve in 
Fig \ref{fig:bigexcitation},) the width becomes $\infty$ and the pulse no longer
exist.  The pulse {\bf l} or {\bf d} can be regained by increasing $u_T$.
However, for large enough $A/a$,   $\Phi(x)$ will become monotonic 
(see Fig.~\ref{fig:newAaexist}) and only {\bf s} exists.

\begin{figure}[htb!]
\centering \epsfig{figure=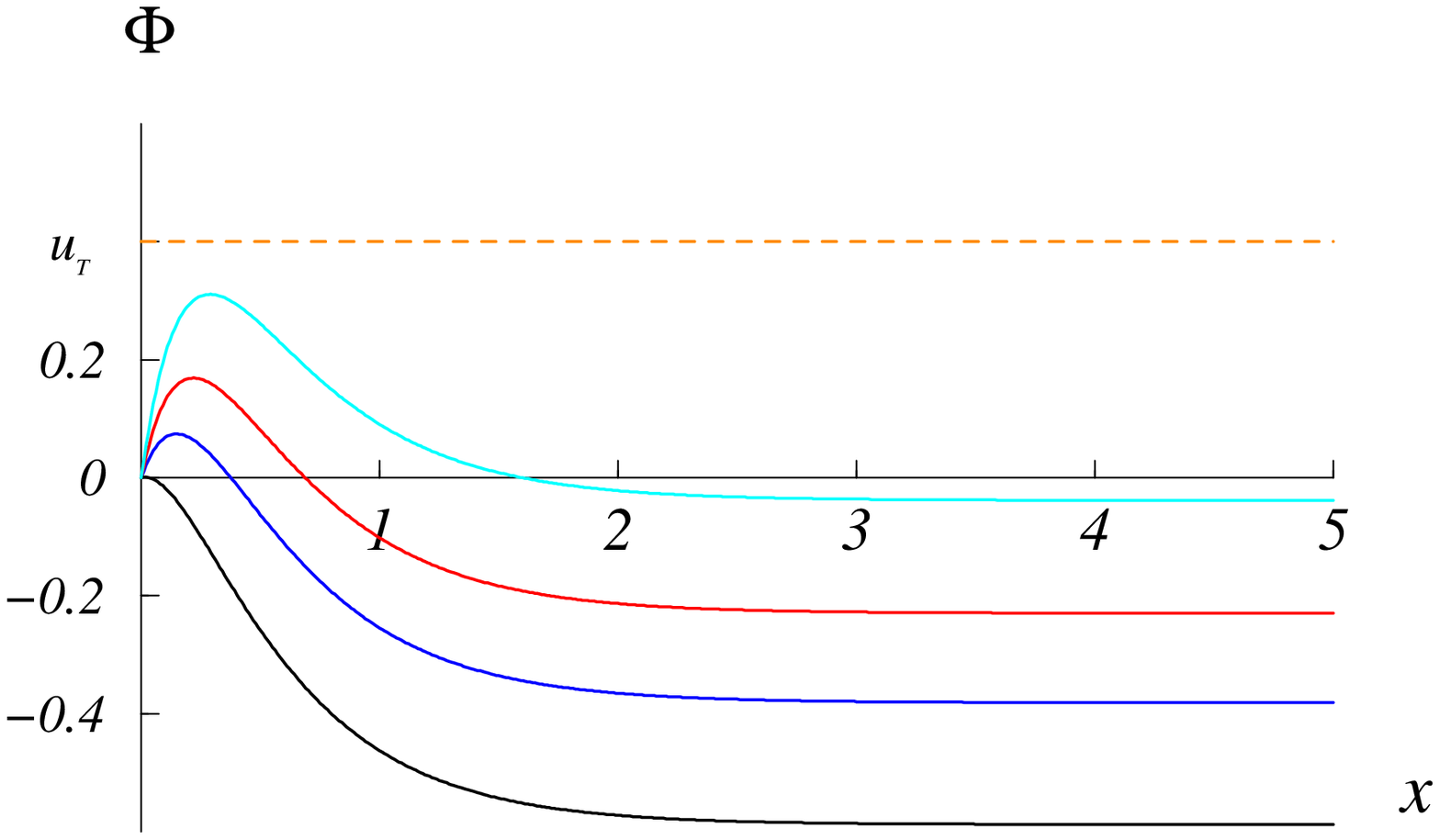, height=1.7in} 
\caption{\small $\Phi(x)$ with inhibition dominating excitation.
  $a=2.6$, $\alpha=0.05$, $u_{\scriptscriptstyle T}=0.400273$ and different
values of $A$: $A=2.5$ (cyan); $A=2$ (blue); $A=1.6$
(red); $A=1.05$ (black). There is neither pulse {\bf s} nor
  pulse {\bf l} for positive $u_{\scriptscriptstyle T}$ when $A/a$ is small 
enough (black color).}
\centering

\label{fig:newaAexist}
\end{figure}

When $A/a<1$, inhibition dominates excitation in the network. For fixed 
$a$ and $\alpha$,  $\Phi(x)$ diminishes as $A$ is decreased (ratio
$A/a$ becomes smaller.) Eventually, the ratio is small enough to make  
$\Phi(x)$ negative (see
Fig.~\ref{fig:newaAexist}), and for $u_T>0$, single-pulses no
longer exist.  Inputs to the neurons in the network never exceed
threshold so the neurons cannot fire.

\subsection{Solutions for complex eigenvalues}
\label{sec:complex} 
As seen in Tables~\ref{tab:eigenvalue1}, \ref{tab:eigenvalue2},
and \ref{tab:eigenvalue3}, the eigenvalues 
$\omega_1$ and $\omega_2$ form a
complex conjugate pair for $\alpha_1<\alpha<\alpha_3$.  Thus as long
as $\alpha_1<\alpha_3$, complex eigenvalues can be found for
arbitrary $a$ and $A$.
%
Suppose $\omega_1=\omega_2^*=p+iq.$ Then
$p=(R^2+S^2)^{\frac{1}{4}}\cos \theta,$ $p=(R^2+S^2)^{\frac{1}{4}}\sin
\theta,$ where $\theta=\frac{1}{2}\arctan \frac{\sqrt{|\Delta|}}{2R}$
for $\alpha \in (\alpha_1, \alpha_2)$ or  
$\theta=\frac{\pi}{2}+\frac{1}{2}\arctan \frac{\sqrt{|\Delta|}}{2R}$
for $\alpha \in (\alpha_2, \alpha_3).$ 
When $\alpha>\alpha_3$, $\Delta>0$, $R<0$ and ${\displaystyle
S=\frac{\sqrt{\Delta}}{2}<|R|}$. The real parts of $\omega_1$ and
$\omega_2$ are both zero and $w_1=iq_1$, $w_2=iq_2,$ where
$q_1=\sqrt{|R+S|}$ and $ q_2=\sqrt{|R-S|}$.

\subsubsection{Construction of a single-pulse with complex eigenvalues}
\begin{figure}[!htb]
\begin{minipage}{2.5in}
\centering
\epsfig{figure=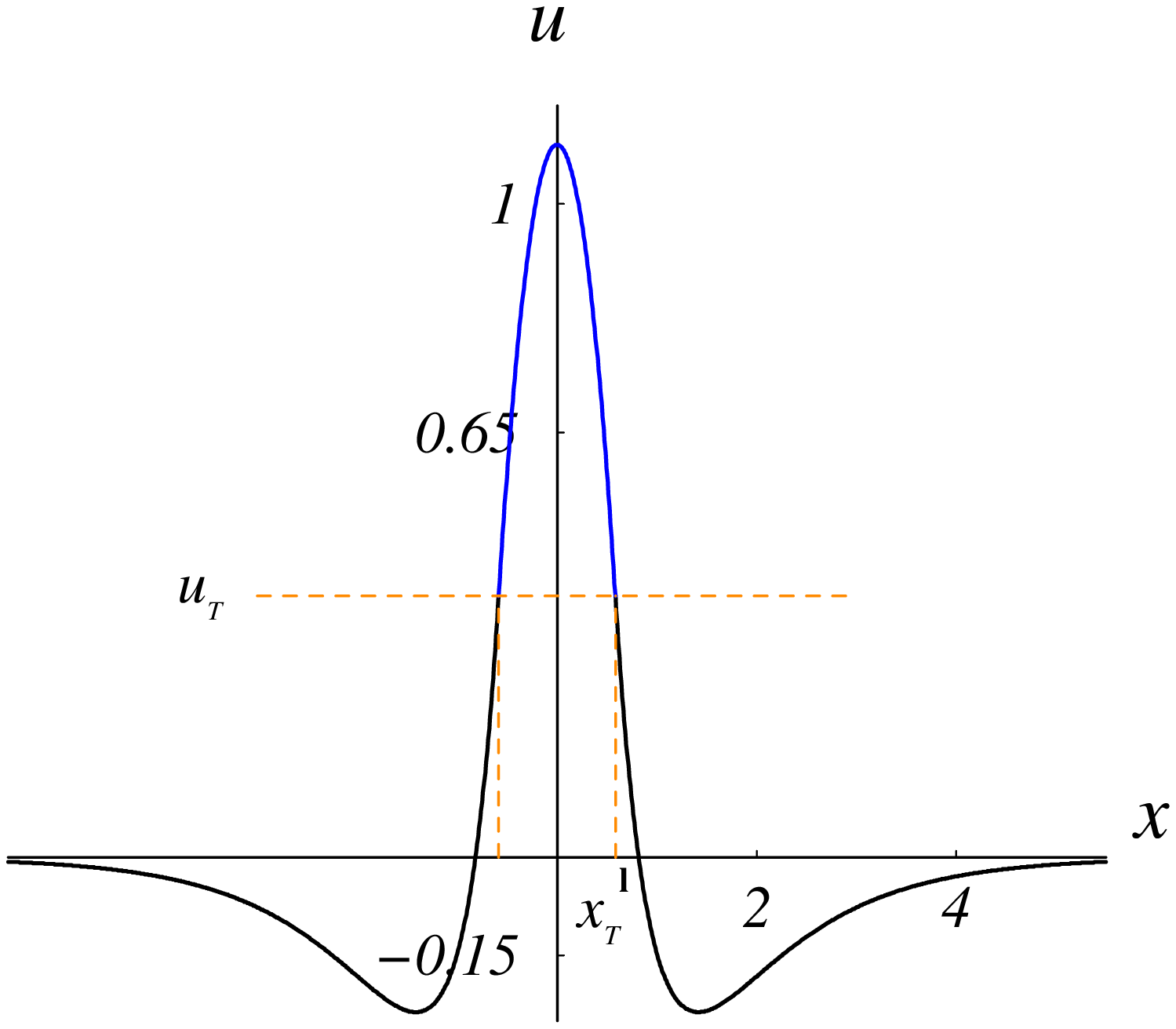, height=2.in}
\end{minipage}
\begin{minipage}{2.5in}
\centering
\epsfig{figure=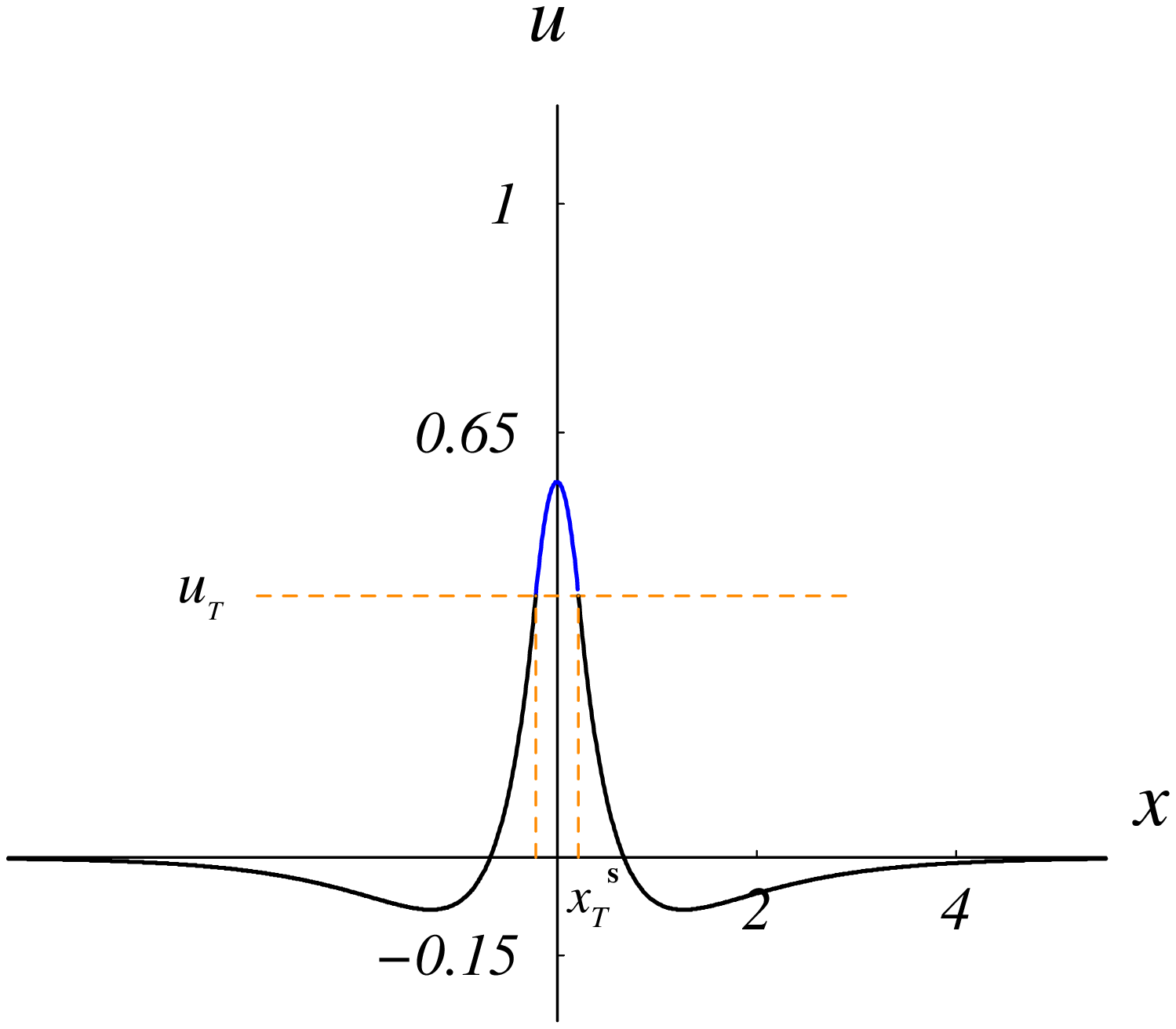, height=2.in}
\end{minipage}
\caption{\small Two single-pulses with $A=2.8$, $a=2.6$, $\alpha=0.6178$,
$u_{\scriptscriptstyle T}=0.3$. (Left) Single-pulse {\bf l} with
$x_{\scriptscriptstyle T}^{\scriptscriptstyle {\bf l}}=0.58384$ and
$u(0)=1.0901$. (Right) Single-pulse {\bf s} with
$x_{\scriptscriptstyle T}^{\scriptscriptstyle \bf{s}}=0.21317$
and $u(0)=0.5744$.}
\centering

\label{fig:comp1smallbump}
\end{figure}

To ensure that $u_{\mathrm I}(x)$ is real, $C$ and $D$ must be
complex. Imposing symmetry, gives $C=D^{*}$. Setting
$C=C_{\scriptscriptstyle R}+iC_{\scriptscriptstyle I}$ implies
$D=C_{\scriptscriptstyle R}-iC_{\scriptscriptstyle I}$. Substituting
$C$, $D$, $\omega_1$ and $\omega_2$ into
(\ref{eq:generalsolregI}) for $u_{\mathrm {I}}(x)$ gives
\begin{eqnarray*}
u_{\mathrm I}(x)=4C_{\scriptscriptstyle
R}\cos(qx)\cosh(px)-4C_{\scriptscriptstyle
I}\sin(qx)\sinh(px)+U_0.
\end{eqnarray*}
For simplicity, we relabel with $C=4C_{\scriptscriptstyle R}$ and
$D=-4C_{\scriptscriptstyle I}$.
When $\omega_1$ and $\omega_2$ are both imaginary, we have
$$
u_{\mathrm I}(x)=C\cos(q_1x_{\scriptscriptstyle
  T})+D\cos(q_2x_{\scriptscriptstyle T})+ U_0.
$$
 Applying the matching conditions (\ref{eq:matchingcondition1})-
(\ref{eq:matchingcondition5}), results in 5 algebraic equations with
unknowns $C,$ $D,$ $E,$ $F$ and $x_{\scriptscriptstyle T}$ which can
 be solved numerically to  obtain the explicit form 
of $u_{\mathrm I}(x)$. The plots of pulses {\bf l} and {\bf s} are
shown in Fig.~\ref{fig:comp1smallbump}.

When $\alpha=\alpha_3$, $R<0$, implying
$\omega_1=\omega_2=\sqrt{R}=i\sqrt{-R}$.
Let $\omega=\sqrt{-R} \in \Re$.
Then 
\begin{eqnarray*}
u_{\mathrm I}(x)=C_1\cos{\omega x}+C_2\sin{\omega x}+D_1x\cos{\omega
  x}+D_2x\sin{\omega x}+U_0. 
\end{eqnarray*}
Since $u_{\mathrm I}(x)=u_{\mathrm I}(-x)$, then  $C_2=D_1=0$ leaving
\begin{eqnarray*}
u_{\mathrm I}(x)=C\cos{\omega x}+Dx\sin{\omega x}+U_0.
\end{eqnarray*}
\begin{figure}[!htb]
\centering
\epsfig{figure=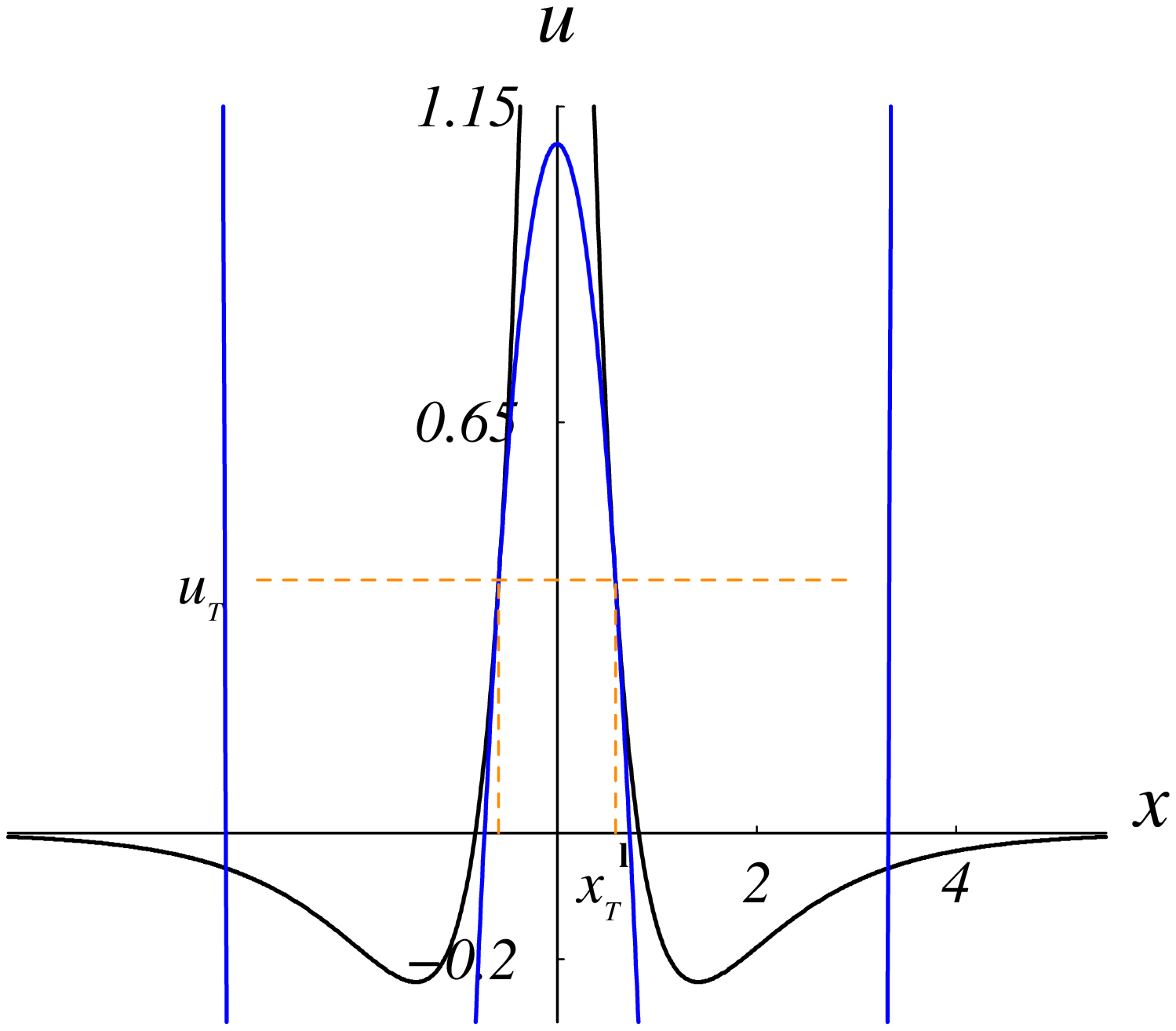, height=2.5in}
\caption{\small Large single-pulse. $A=2.8$, $a=2.6$, $\alpha=0.6178$,
$u_{\scriptscriptstyle T}=0.3,$ $x_{\scriptscriptstyle
    T}^{\scriptscriptstyle {\bf l}}=0.58384$, and $u(0)=1.0901$.} 
\label{fig:comp1bumpmatch}
\end{figure}

\subsubsection{Finding solutions for complex eigenvalues using the existence function}
\label{sec:comp1existencefunction}
\begin{figure}[!htb]
\centering \epsfig{figure=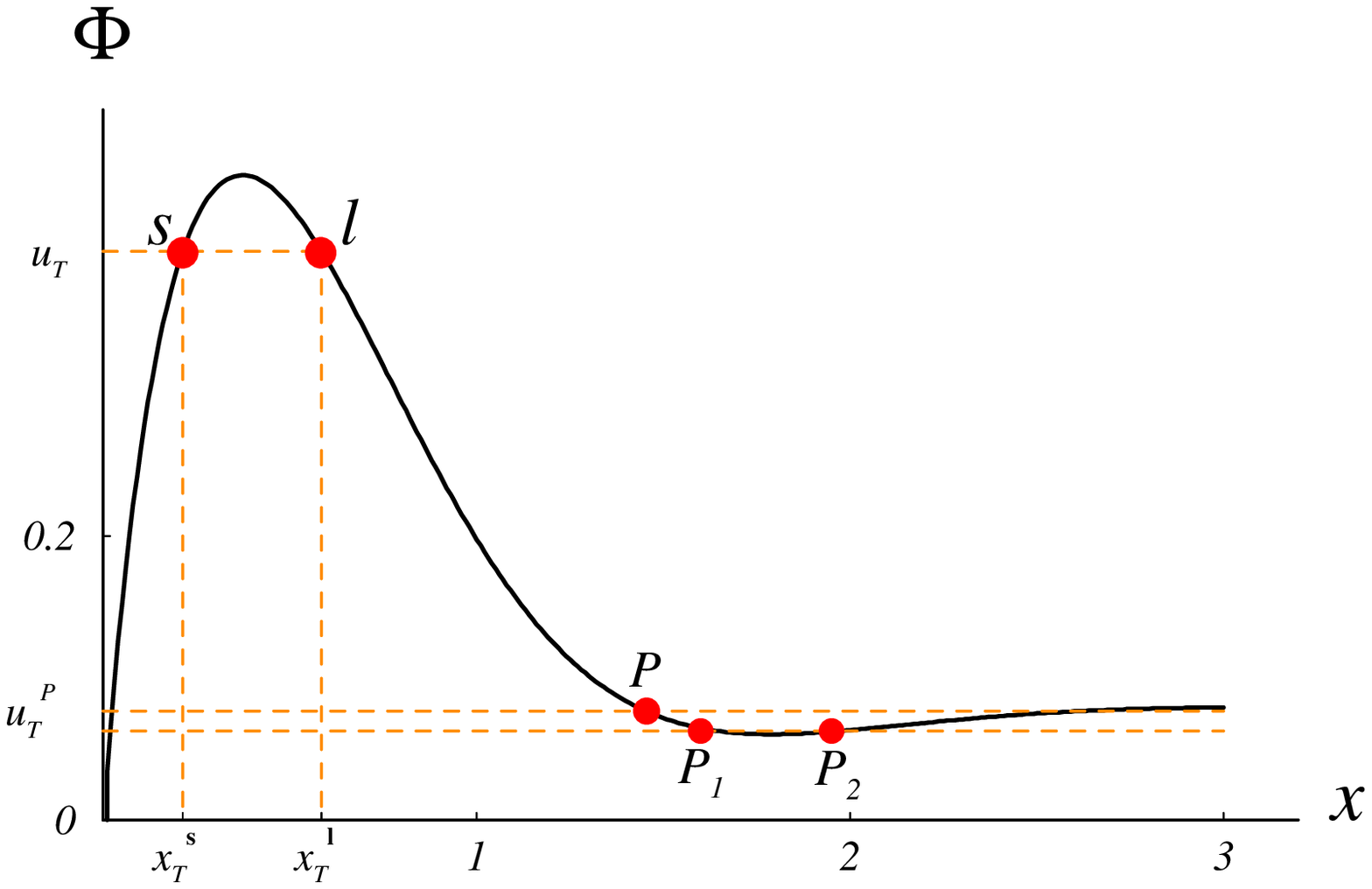, height=2.6in}
\caption{\small Existence function $\Phi(x)$. $\alpha=0.6178$,
$A=2.8$, $a=2.6$. At $u_{\scriptscriptstyle T}=0.400273$, $\Phi(x)$
shows that there is a single-pulse {\bf l} which is wider and has
width $x_{\scriptscriptstyle T}^{\scriptscriptstyle {\bf l}}=0.58385$;
the second single-pulse {\bf s} is narrower and has width
$x_{\scriptscriptstyle T}^{\scriptscriptstyle {\bf s}}=0.21317$. As we
increase $u_{\scriptscriptstyle T}$ to the maximum of $\Phi(x)$, pulse
{\bf s} and {\bf l} annihilate in a saddle-node bifurcation. At $P$,
$u_{\scriptscriptstyle 
T}^{\scriptscriptstyle P}=0.0767$, $x_{\scriptscriptstyle
T}^{\scriptscriptstyle P}=1.454$, $u''(0)=0$. At both $P_1$ and $P_2$,
$u_T=0.063$, $u''(0)>0$, and the widths are $1.6$ and $1.9$
respectively. See Fig.~\ref{fig:comp1dimple}.} \centering
\label{fig:comp1exist}
\end{figure}

\begin{figure}[!htb]
\begin{minipage}{2.5in}
\centering \epsfig{figure=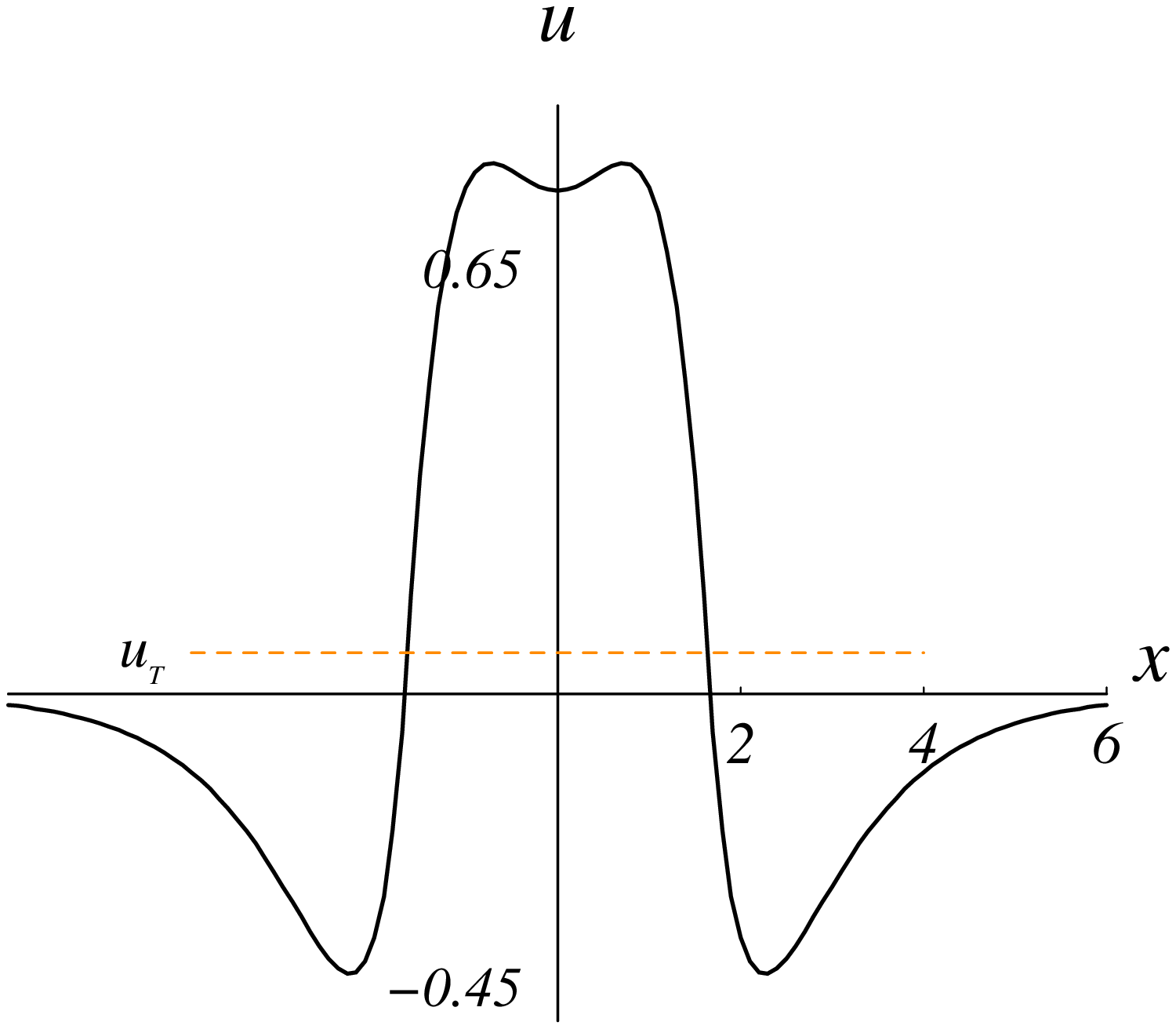, height=2in}
\end{minipage}
\begin{minipage}{2.5in}
\centering \epsfig{figure=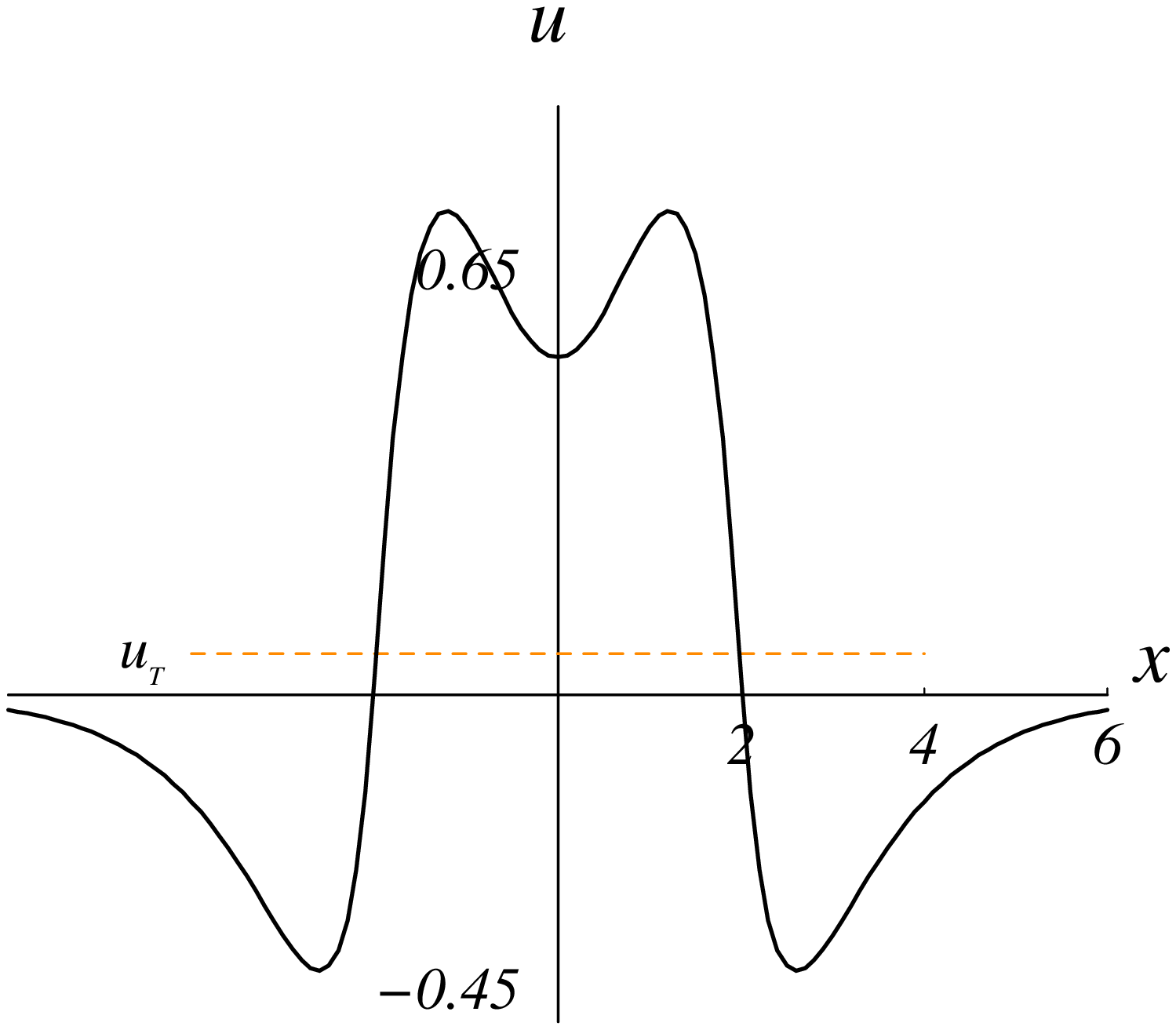, height=2in}
\end{minipage}
\caption{\small Dimple-pulses with parameters $A=2.8$, $a=2.6$,
$\alpha=0.6178$, $u_{\scriptscriptstyle T}=0.063$. (Left) Dimple-pulse
at $P_1$ with $x_{\scriptscriptstyle T}^{\scriptscriptstyle {\bf
d}}=1.6$. (Right) Dimple-pulse at $P_2$ with $x_{\scriptscriptstyle
T}^{\scriptscriptstyle {\bf d}}=1.9$.} \centering
\label{fig:comp1dimple}
\end{figure}

\begin{figure}[!htb]
\centering
\includegraphics[width=4.5in]{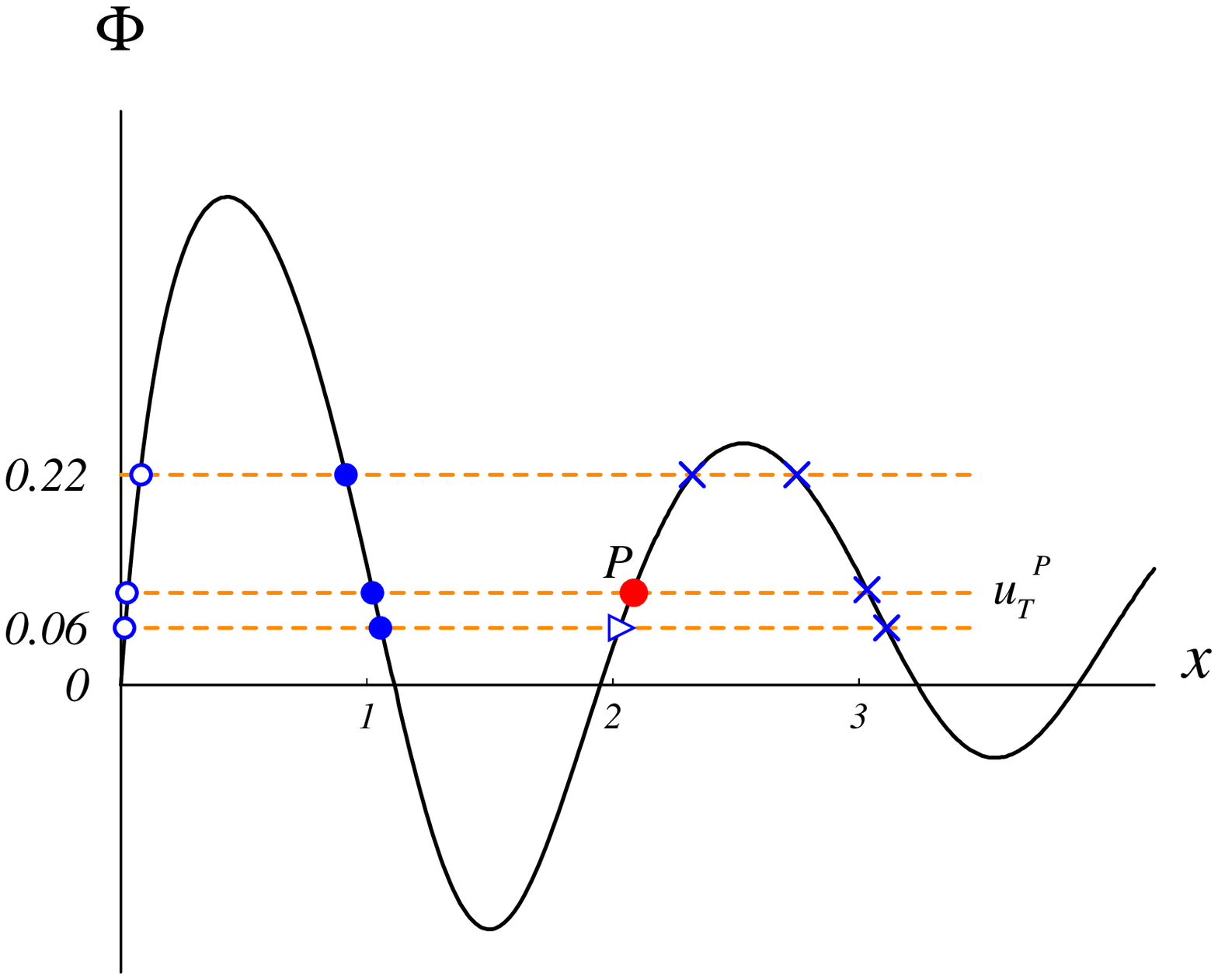}
\caption{\small Existence function $\Phi(x)$ with imaginary
  $\omega_1=\omega_2$ for $A= 2.8$, $a=2.6$, $\alpha=\alpha_3$,
  $u_{\scriptscriptstyle T}^P=0.0967003$. The empty circles are
  small single-pulses. The solid circles are large
  single-pulses. The triangle is a dimple-pulse. Point $P$ is where
  the dimple-pulse (figure \ref{fig:dimple2double}) breaks into a
  double-pulse. The $\times$s are not valid solutions.} \centering
\label{fig:eqeivaexist}
\end{figure}

The existence function $\Phi(x)$ with complex $\omega_1$ and
$\omega_2$ (Figs.~\ref{fig:comp1exist} and \ref{fig:eqeivaexist}) can
be obtained using methods similar to those in
Sec.~\ref{sec:realexistencefunction}.  The main difference with the
real case is that $\Phi(x)$ for complex eigenvalues can oscillate as
seen in Fig.~\ref{fig:comp1exist}.  After the first local minimum
between $P_1$ and $P_2$, $\Phi(x)$ will approach a constant with
decaying oscillations for increasing $x$.  Additionally, if the
threshold is between the first local minimum and the next local
maximum, there exist more than two pulse
solutions. Figure~\ref{fig:comp1dimple} shows an example where there
is a small single-pulse and two dimple-pulses.  There is never
more than two coexisting pulses for real $\omega_1$ and $\omega_2$,
which includes the Amari case, because the existence function $\Phi(x)$
does not oscillate.  It is important to note that satisfying the
existence condition $\Phi(x)=u_T$ is a necessary but not a sufficient
condition for a pulse solution.  It is possible to satisfy the
matching conditions and not be a pulse.  Thus although
in principle, an infinite number of pulses could exist, in our
experience, we find that most of the larger $x$ solutions are not
pulses.  As will be shown in the accompanying paper~\cite{Guo3}, pulse
{\bf s} is unstable and pulse {\bf l} is stable.  If there are three
pulses, the largest third pulse which can be either a single pulse or
a dimple pulse is unstable. 


Oscillations in $\Phi(x)$ also exist when the eigenvalues are pure
imaginary.  As before, more than two pulses, including dimple-pulses,
can coexist (see Fig.~\ref{fig:eqeivaexist}) depending on the
threshold $u_{\scriptscriptstyle T}$.  Fig.~\ref{fig:eqeivaexist}
shows a special case of a dimple-pulse where the dimple minimum
reaches the threshold. If the minimum drops below the threshold, the
dimple-pulse breaks into two disjointed single-pulses or a
double-pulse.  This double-pulse is not a valid solution
because it violates the assumptions of the equations from which the
solution was derived.  However, double pulses can exist and we show
this using a separate formalism in Sec.~\ref{sec:doublepulse}. 

\begin{figure}[!htb]
\centering
\includegraphics[width=2.5in]{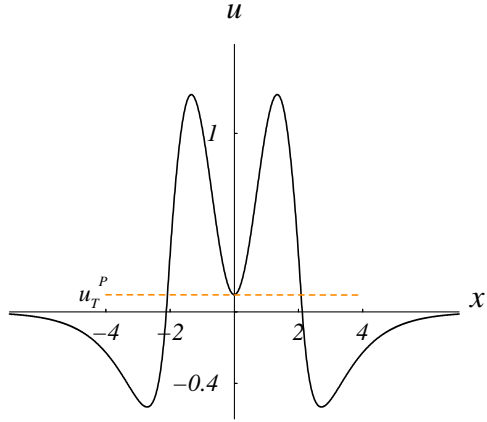}
\caption{\small The transition from a dimple-pulse to a double-pulse at $P$
with $u_{\scriptscriptstyle T}^P=0.0967003$, and $u(0)=0.0967003$. }
\centering
\label{fig:dimple2double}
\end{figure}

\subsubsection{Blow-up for large $\alpha$}
\label{sec:comp3exist}

For large enough $\alpha$, the large pulse {\bf l} blows up at a
critical value 
$\alpha^0$ and does not exist for $\alpha\ge \alpha_0$.  The blow-up
occurs in the regime where both $\omega_1$ and 
$\omega_2$ are imaginary. Thus the height of pulse {\bf l} is
\begin{equation}
u(0)  =  C+D+\frac{2(A - a)(\beta - \alpha u_{\scriptscriptstyle
T})}{a - 2\alpha(A - a)},
\end{equation}
which can be expressed as
\begin{equation}
u(0) = \frac{\left| \begin{array}{cccc}
(m_1-m_2) & m_0 & m_3 & m_4
\end{array}\right|}{\left|\begin{array}{cccc} m_1 & m_2 & m_3 & m_4
  \end{array}\right|}, 
\label{numerator}
\end{equation}
where the coefficient factors $m_1,$ $m_2,$ $m_3,$ $m_4$ for $C,$ $D,$
$E,$ $F$ and $m_0$ are defined in section
\ref{sec:realexistencefunction}.  The blow-up occurs because the
denominator of (\ref{numerator}) goes to zero at $\alpha=\alpha^0$
while the numerator remains finite, sending
the height $u(0)$ of the large pulse to infinity. 

\begin{figure}[!htb]
\begin{minipage}{2.5in}
\centering \epsfig{figure=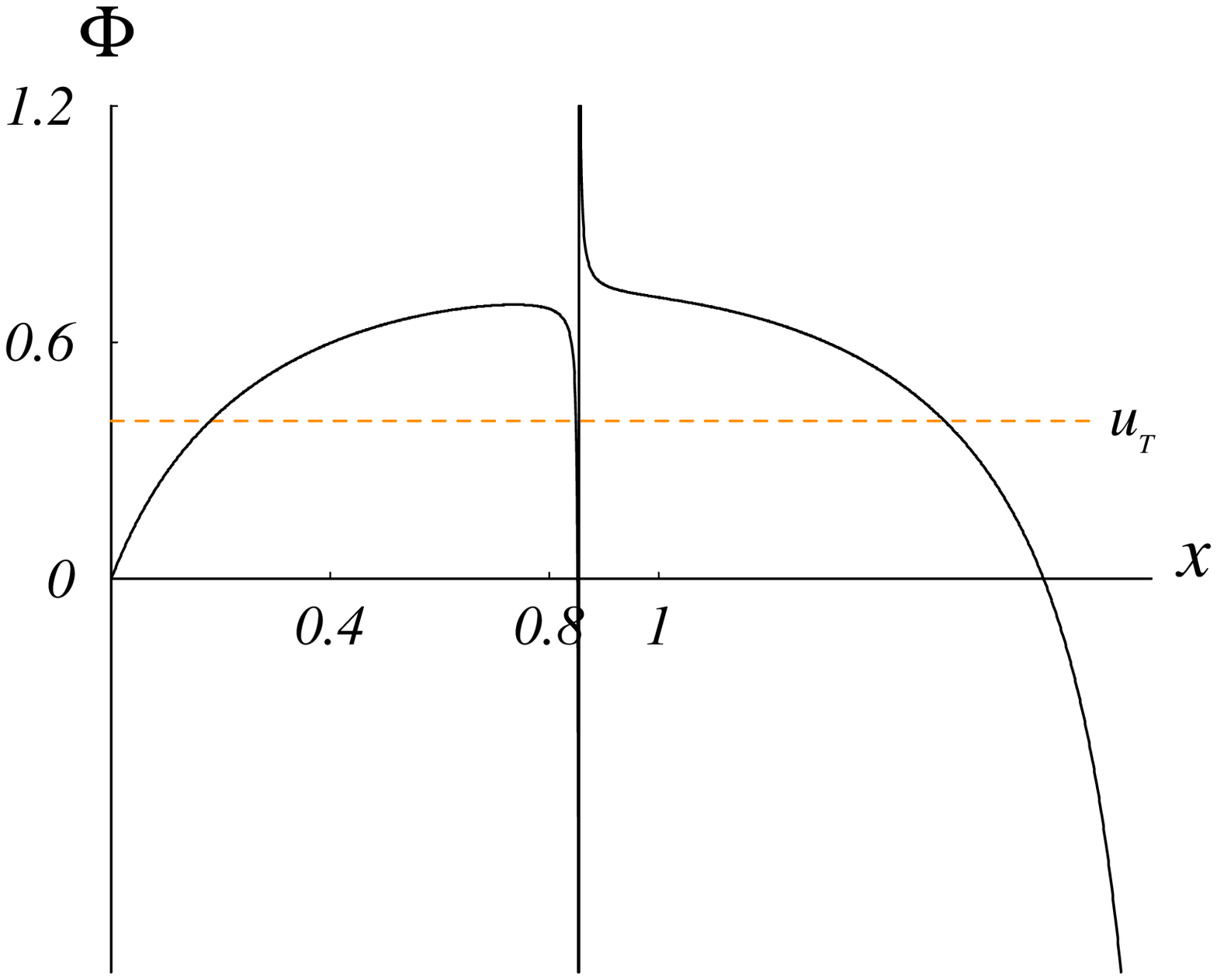, height=1.8in}
\end{minipage}
\begin{minipage}{2.5in}
\centering \epsfig{figure=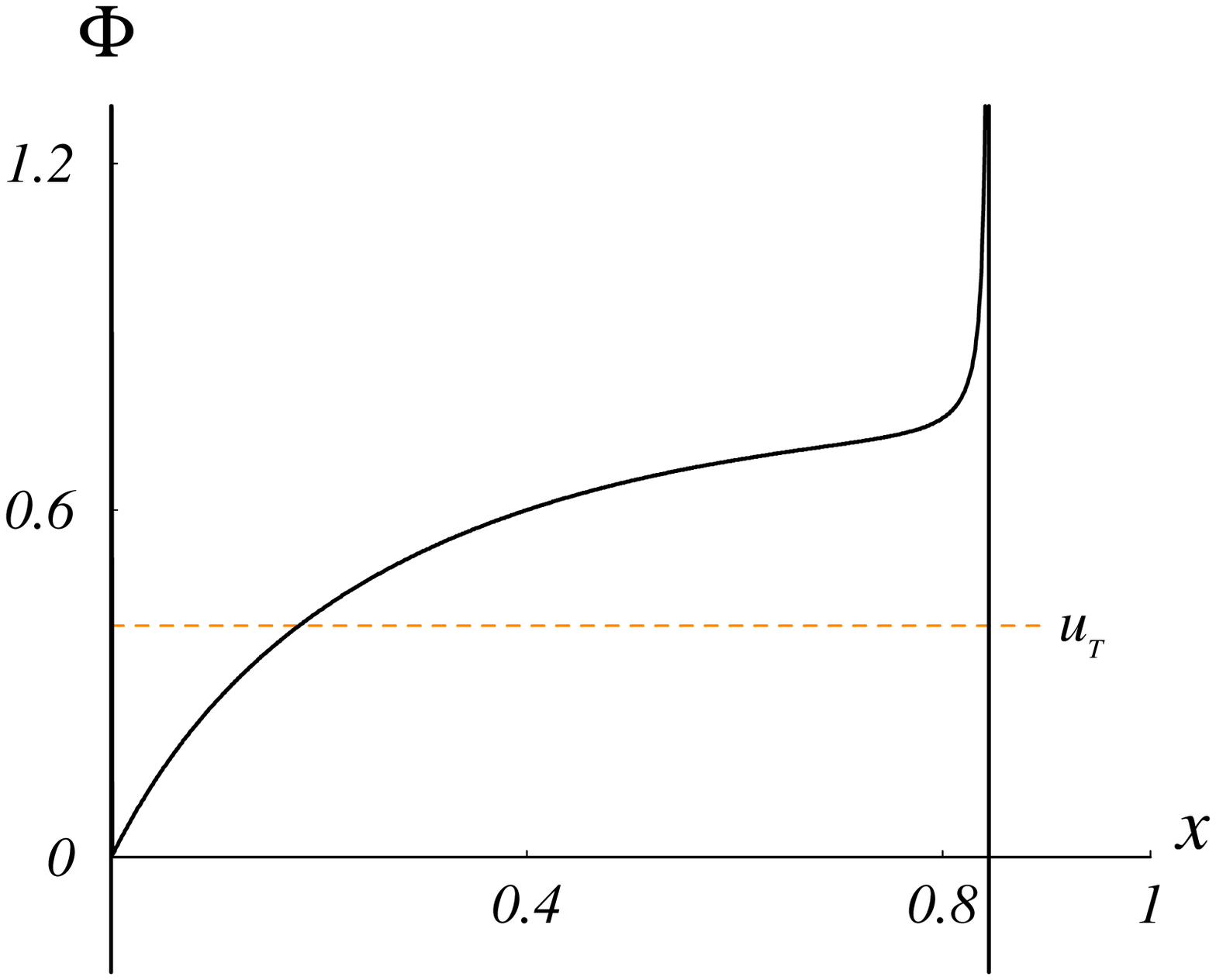, height=1.8in}
\end{minipage}
\caption{\small Existence function $\Phi$ for imaginary
$\omega_{1,2}$ with $A=2.8$, $a=2.6$, $u_{\scriptscriptstyle
T}=0.400273$. (Left) $\alpha=1.4$. There is a single-pulse {\bf l},
and $x_{\scriptscriptstyle T}^{\scriptscriptstyle {\bf
l}}=0.8491539857774331$, height=$u(0)=146.2227855915919$, which is big
because $\alpha=1.4$ is close to $\alpha^0$ where $DET=0$. (Right)
$\alpha=1.41 > \alpha^0$. Single-pulse {\bf l} no longer exists. The
vertical line in both pictures is where $\Phi(x)$ blows up.} 
\label{fig:comp3exist}
\end{figure}

We can also see the loss of {\bf l} in the existence function:
$$\Phi(x)=E e^{-ax}+F e^{-x}=\frac{\left|\begin{array}{cccc} m_1 & m_2
& m_0 & (m_3-m_4) \end{array}\right|}{\left|\begin{array}{cccc} m_1 &
m_2 & m_3 & m_4 \end{array}\right |}.$$
Figure~\ref{fig:comp3exist} shows that
when $\alpha \geq \alpha^0$, there is always the small single-pulse
${\bf s}$ but no large single-pulse {\bf l}. A third solution (the
third intersection of $u_{\scriptscriptstyle T}$ and $\Phi(x)$) could
also exist but we have not examined this solution.
As $\alpha$ increases, the height of {\bf l}
becomes very large but the width of the large pulse remains finite. This
can be observed both from the existence function $\Phi(x)$ and the
continuation plot (Fig \ref{fig:heightautopatch0})
in Sec.~\ref{sec:autoandglobal}. 

\section{Continuation in parameter space}
\label{sec:autoandglobal}

One of our original goals was to understand how the shape of
stationary pulses and their corresponding firing rates change as the
parameters of synaptic connectivity and gain are changed.  Here we
give a global picture in the parameter space of $u_T$, $a$, $A$, and
$\alpha$.   A difficulty in this undertaking is that
as the parameters are altered, the eigenvalue structure will make
abrupt transitions.  Hence, one must keep track of the eigenvalues and
switch the form of the solutions when appropriate to construct a
global picture. 

\begin{figure}[!htb]
\centering \epsfig{figure=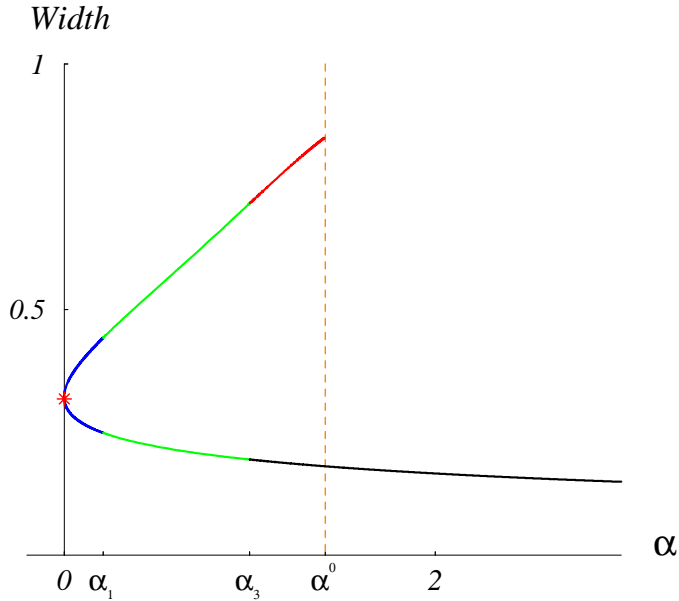, height=3in}
\caption{\small Width  of single-pulse {\bf l} (upper branch) and {\bf
s} (lower branch) for $a=2.6$, $A=2.8$, and $u_{\scriptscriptstyle
T}=0.400273$. For $\alpha \in [\alpha^*,\alpha^0)$, there 
are two single-pulses. For $\alpha \in [\alpha^0, \infty)$, there is only
one single-pulse solution. At $\alpha=0$ there is a saddle-node
bifurcation where the large
single-pulse {\bf l} and the small single-pulse {\bf s} arise.
At $\alpha^0$, the large single-pulse {\bf l} blows up.} 
\label{fig:widthautopatch0}
\end{figure}

\begin{figure}[!htb]
\centering \epsfig{figure=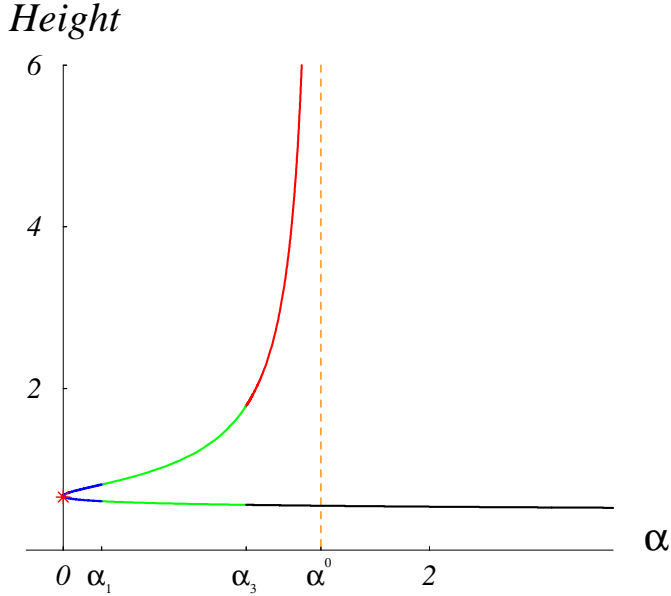, height=3in}
\caption{\small Height of single-pulse {\bf l} (upper branch) and
{\bf s} (lower branch) for the same conditions as
Fig.~\ref{fig:widthautopatch0}.
} \centering
\label{fig:heightautopatch0}
\end{figure}

As we saw before, the small and large pulses arise out of a
saddle-node bifurcation.  This gives a minimal condition for when
pulses can exist.  
In Figs.~\ref{fig:widthautopatch0} and \ref{fig:heightautopatch0}, we
show the large pulse {\bf l}
and the small pulse {\bf s} arising from a saddle node bifurcation as
$\alpha$ is increased for fixed $a$ and $A$.  
We have set the threshold to
\begin{eqnarray*}
u_{\scriptscriptstyle T}^{0}=\int_{0}^{\ln A/(a-1)}w(x)dx
\end{eqnarray*}
so that the saddle node is exactly at $\alpha=0$. 
Note that at the saddle node bifurcation, the pulse arises with nonzero
height and width.

We can now track the location of the saddle node and
the maximum firing rate of the pulse at the saddle node in
parameter space. We reduce the four dimensional parameter space by
projecting to the space $(\alpha$, $a/A$, $u_{\scriptscriptstyle T})$.  The
saddle-node location can be found by setting the threshold $u_T$ to
the value of the first local maximum of $\Phi(x)$.  This gives an
upper bound for allowable thresholds of the gain function to support a
pulse solution.  As long as $u_T$ is below this upper bound and
positive, a single-pulse can exist.

\begin{figure}[t]
\centering \epsfig{figure=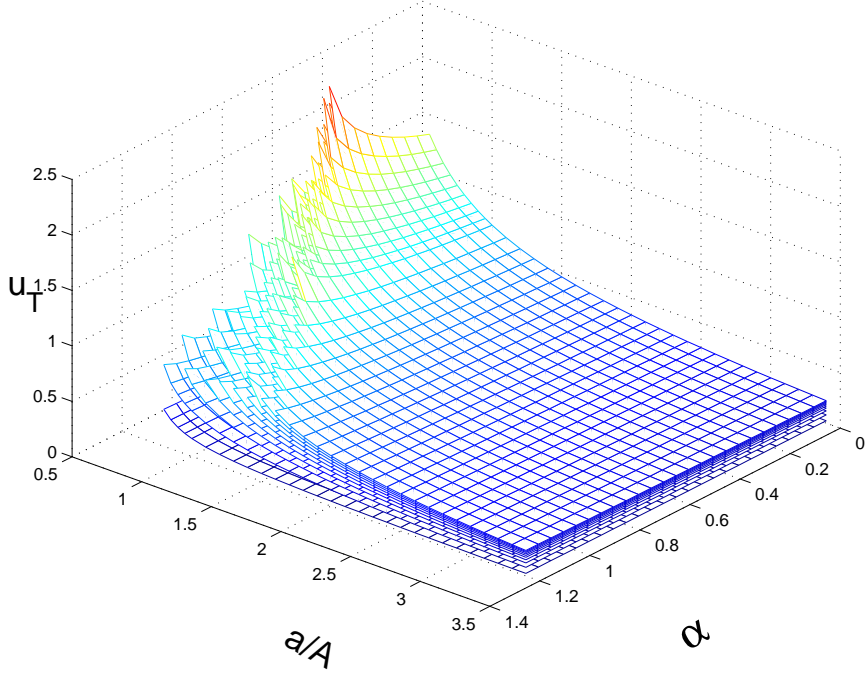, height=3.6in} \caption{\small Surface
  plot of saddle node point in parameter space of $(a/A,\alpha,u_T)$.
  The separate leaves correspond to values of $A$ ranging from [*] to
  [*].} 
\centering
\label{fig:globalut}
\end{figure}

\begin{figure}[t]
\centering \epsfig{figure=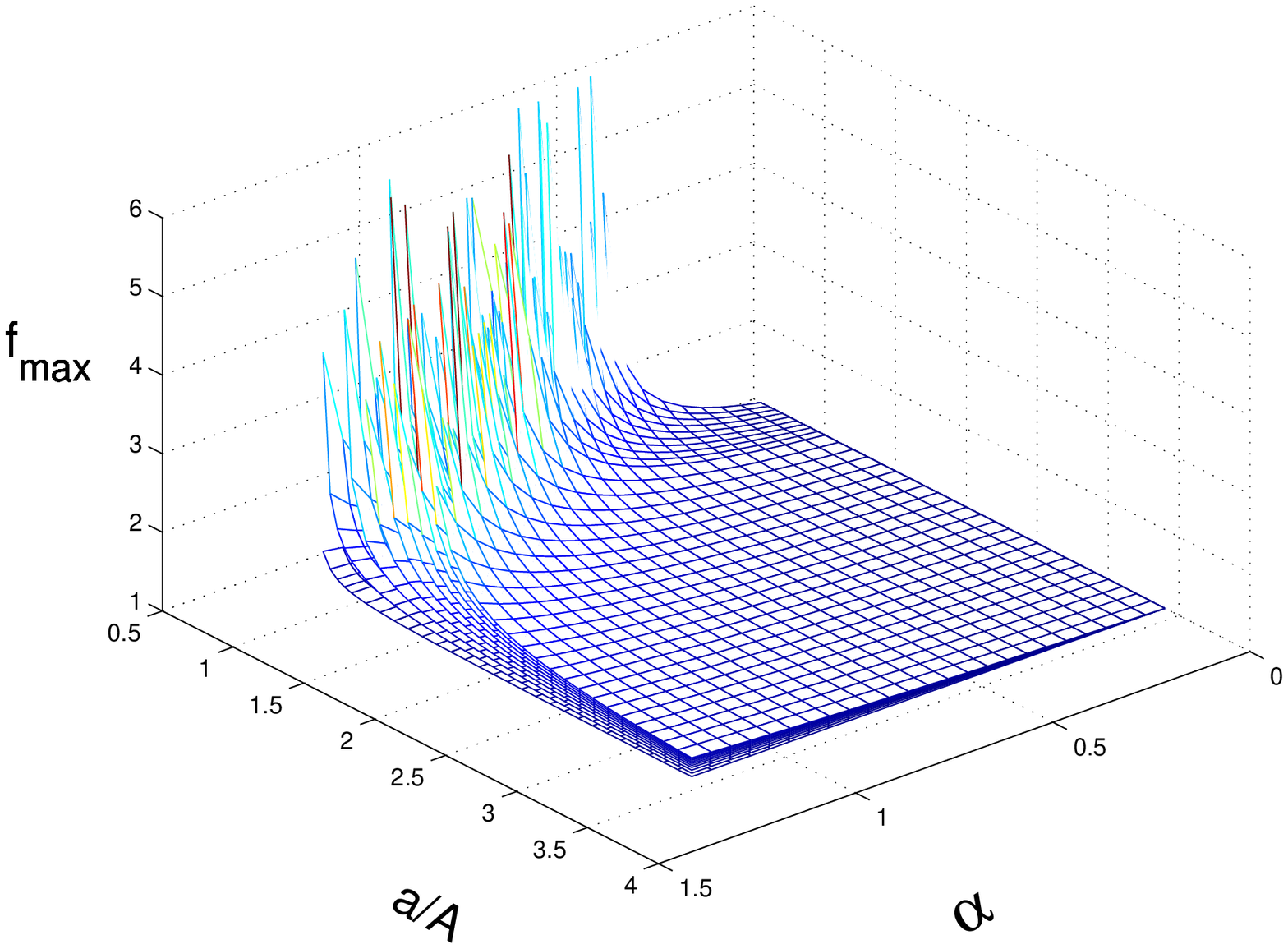, height=3.6in} \caption{\small
  Maximum point of firing rate of the single-pulse at the saddle node
  for the conditions as Fig.~\ref{fig:globalut}.} 
\centering
\label{fig:globalgain}
\end{figure}

We first set $A=1.5$ and vary the ratio $a/A$ and $\alpha$ to
identify the saddle node threshold $u_{\scriptscriptstyle T}$. We then
calculate the maximum firing rate $f_{max}$ of the single-pulse
solution at this threshold which creates a two dimensional surface in
the space of $(a/A, \alpha, u_T)$.  We increment $A$ in steps of
$1$ and create a set of surfaces.  The surface plots of
$u_{\scriptscriptstyle T}$ and $f_{max}$ vs $a/A$ and $\alpha$ are
shown in Figs.~\ref{fig:globalut} and \ref{fig:globalgain}.
Single-pulse solutions exist below a given surface (with $u_{\scriptscriptstyle T}>0$). Depending on the parameters, solutions could include one single-pulse
${\bf s}$, a coexistence of single-pulses {\bf s } and 
{\bf l} (or a dimple-pulse {\bf d} but in a smaller global range), or
coexistence of more than two pulses.

When excitation dominates inhibition (i.e. $a/A$ is small) the
single-pulse solution can blow up as mentioned in
Sec.~\ref{sec:comp3exist}.  We note that the crucial parameter for
maintaining low firing rates is for inhibition to dominate excitation
(i.e. $a/A$ to be large).  Even for the balanced case of $a=A$, for
gain slope $\alpha$  beyond unity, the firing rate rises
dramatically.   This is in correspondence with observations of
numerical simulations \cite{Wang1, Wang}.

\section{Construction of double-pulse solutions}
\label{sec:doublepulse}
\begin{figure}[!htb]
\centering
\includegraphics[height=2.4in, width=5in]{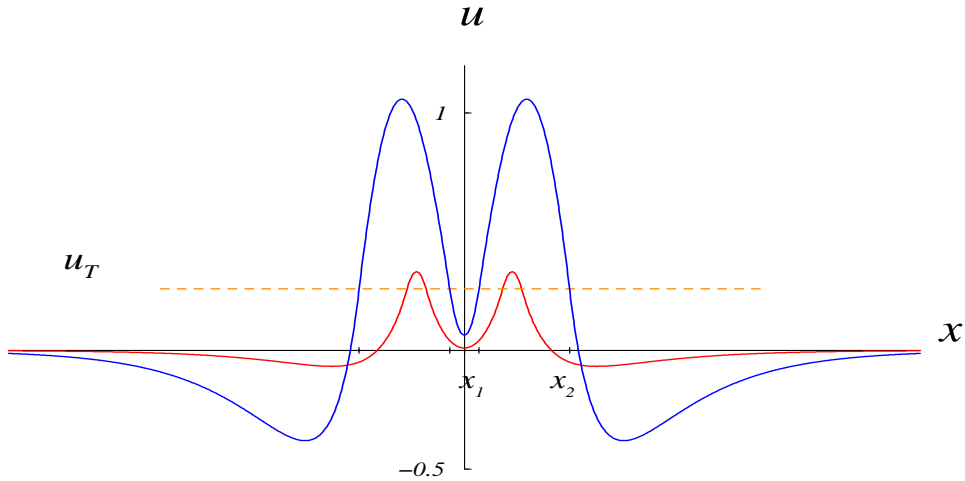}
\caption{\small Two coexisting double-pulses. $A=2.8$, $a=2.6$,
  $\alpha=0.98$, $u_{\scriptscriptstyle T}=0.26$. For the large
  double-pulse (blue): $x_1=0.19266$, $x_2=1.38376$. For the small
  double-pulse (red): $x_1=0.50582$, $x_2=0.752788$.} 
\label{fig:comp2doublebump}
\end{figure}

The neural network equation 
(\ref{eq:differentiate}) can also support double-pulse or even multiple-pulse solutions~\cite{Guo1, Troy2}. 
Double-pulses are solutions that have two
disjoint open and finite intervals for which the synaptic input $u(x)$
is above threshold.
\begin{definition}
\label{def:doublepulse}
{\bf Double-pulse solution}:
A solution $u(x)$ of (\ref{eq:integral_equation2}) is called a double-pulse or
a 2-pulse if there are $x_1>0$ and $x_2>0$ such that
$$u(x) \left \{ \begin{array}{ll}
>u_{\scriptscriptstyle T} & \mbox{$\mathrm{if}$ $x\in(x_1,x_2) \cup
  (-x_2, -x_1)$, $x_{1,2}>0$}\\ 
=u_{\scriptscriptstyle T} & \mbox{$\mathrm{if}$ $x=-x_2,-x_1, x_1, x_2$} \\
<u_{\scriptscriptstyle T} &  \mbox{$\mathrm{otherwise}$}
\end{array} \right.$$ with
\begin{eqnarray*}
(u,u',u'',u''') \rightarrow (0,0,0,0)
\end{eqnarray*}
exponentially fast as $x \rightarrow \pm \infty.$   $u$ and $u'$ are
bounded and continuous on $R$.  $u'',$ $u'''$ and $u''''$ are
continuous everywhere for $x \in \Re$ except  
$x=\pm x_{1,2}$ and bounded everywhere on $R$. $u(x)$ is symmetric
with $u''(0)>0;$ $u(0)$ is the 
minimum  between $-x_1$ and $x_1$ $\mathrm{( Fig.}$
$\mathrm{\ref{fig:comp2doublebump})}$.  
\end{definition}

The approach to find and construct double-pulse solutions is similar to that for
single-pulses.  The connection function is (\ref{eq:connection}) and
the gain function is (\ref{eq:firingrate}).    Laing and
Troy~\cite{Troy2} found that double pulses can exist for the Amari
case ($\alpha=0$).  
However, for the exponential connection function
(\ref{eq:connection}), the double pulses are unstable.  
Coombes {\it et al.}~\cite{Coombes} found that
double and higher number
multiple-pulse solutions could exist in a network with a saturating
sigmoidal gain function.
As in the
single-pulse case, a fourth order ODE on $x \in (-\infty, \infty)$ for
double-pulses can be derived: 
\begin{eqnarray}
\label{eq:4thODE}
\label{eq:double4thode}
\lefteqn{u^{''''}-(a^2+1)u''+a^2u=} \\ \nonumber
&  &
2a(A-a)f\left[u(x)\right]+2(aA-1)
\left\{f[u(x_2)]\Delta_2'(x)-f[u(x_1)]\Delta_1'(x)\right\}-   
\\ \nonumber & &  \mbox{} 2(aA-1)\left\{f'[u(x_1)]u'(x_1)
\Delta_1(x)+f'[u(x_1)]u'(x_1)\Delta_2(x)\right\}- \\  \nonumber & &
\mbox{} 2(aA-1) \frac{d^2f\left[u(x)\right]}{dx^2}
\end{eqnarray}
where
\begin{eqnarray*}
\Delta_1(x) & = & \delta(x-x_1)+\delta(x+x_1) \\
\Delta_2(x) & = & \delta(x-x_2)+\delta(x+x_2) \\
\Delta_1'(x) & = & \delta'(x-x_1)-\delta'(x+x_1) \\
\Delta_2'(x) & = & \delta'(x-x_2)-\delta'(x+x_2) 
\end{eqnarray*}
Double-pulses can be constructed using ODE (\ref{eq:double4thode}) and 
the following set of matching conditions
(\ref{eq:double1})-(\ref{eq:double10}) at both $x_1$ and $x_2$: 
\begin{eqnarray}
\label{eq:double1}
u_I(x_1) & = & u_{\scriptscriptstyle T} \\
u_{II}(x_1) & = & u_{\scriptscriptstyle T} \\
u_{II}(x_2) & = & u_{\scriptscriptstyle T} \\
u_{III}(x_2) & = & u_{\scriptscriptstyle T} \\
u_I'(x_1) & = & u_{II}'(x_1) \\
u_{I}''(x_1) & = & u_{II}''(x_1)+2(aA-1)f(u(x_1)) \\
u_{I}'''(x_1) & = & u_{II}'''(x_1)+2(aA-1)f'(u(x_1))u'(x_1) \\
u_{II}'(x_2) & = & u_{III}'(x_2) \\
u_{II}''(x_2) & = & u_{III}''(x_2)-2(aA-1)f(u(x_2)) \\
u_{II}'''(x_2) & = & u_{III}'''(x_2)-2(aA-1)f'(u(x_2))u'(x_2)
\label{eq:double10}
\end{eqnarray}

\begin{figure}[!htb]
\centering
\includegraphics[height=2.4in, width=5in]{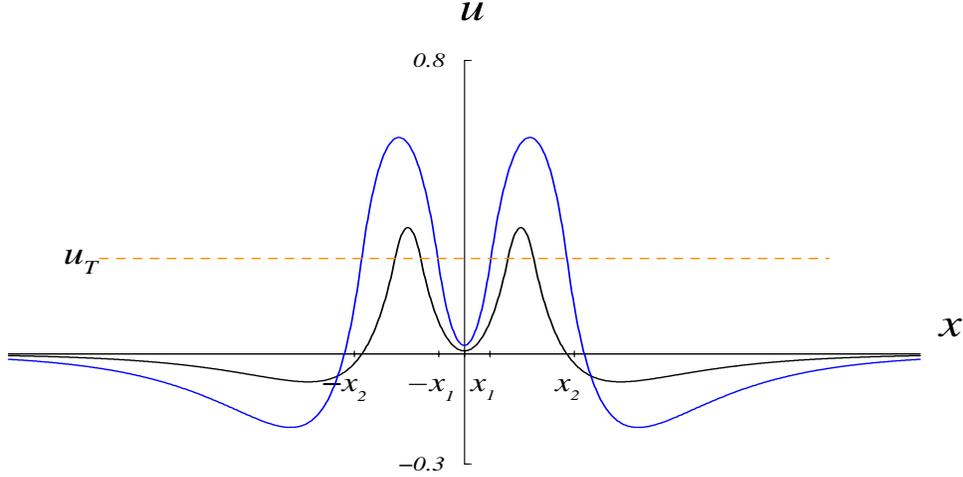}
\caption{\small Double-pulse for Amari case in which
  $\alpha=0$. $A=2.8$, $a=2.6$, $\alpha=0$, $u_{\scriptscriptstyle
    T}=0.26$. For the large double-pulse(blue): $x_1=0.279525$,
  $x_2=1.20521$. For the small double-pulse (black): $x_1=0.49626$,
  $x_2=0.766206$. } 
\label{fig:amaridoublebump}
\end{figure}

In the Amari case ($\alpha=0$), for the same set of values of $a,$ $A,$ 
$u_{\scriptscriptstyle T}$ we find that there are two coexisting
double-pulse solutions. The existence  
conditions are 
\begin{eqnarray}
f_1(x_1,x_2)=u(x_1) & = &
\int^{-x_1}_{-x_2}w(x_1-y)dy+\int^{x_2}_{x_1}w(x_1-y)dy =
u_{\scriptscriptstyle T} \\ 
f_2(x_1,x_2)=u(x_2) & = &
\int^{-x_1}_{-x_2}w(x_2-y)dy+\int^{x_2}_{x_1}w(x_2-y)dy =
u_{\scriptscriptstyle T} 
\end{eqnarray}
Both $x_3=f_1(x_1,x_2)$ and $x_3=f_2(x_1,x_2)$ form two dimensional
surfaces in the three dimensional space $(x_1,x_2,x_3)$.  The two
surfaces intersect in a convex up space curve.  The intersection of
this space curve with the plane 
$x_3=u_{\scriptscriptstyle T}$ are widths of candidate double pulse solutions.
Since the surfaces do not oscillate for the Amari case there are only
two intersection points. These two points give two double pulse
solutions which are shown in  
Fig. \ref{fig:amaridoublebump}.  For the general $\alpha>0$ case we, have
 found two 
coexisting double-pulses as shown in Fig
 \ref{fig:comp2doublebump}. The coexistence of more than two 
double-pulses for $\alpha >0$ remains an open problem.

In the accompanying paper~\cite{Guo3} we confirm the Laing and Troy
finding~\cite{Troy2} that the double pulses in the Amari case are
unstable.

\section{Discussion}
\label{sec:conclusions}

In this paper, we consider a population neural network model of the form
(\ref{eq:differentiate}) with a nonsaturating gain function of the
form (\ref{eq:piecewiseligain}). We show the existence of stationary
solutions that satisfy the equilibrium equation
(\ref{eq:integral_equation2}) by explicitly constructing single-pulse
solutions for a specific synaptic connection
function (\ref{eq:connection}). The strategy was to  
convert the integral equation (\ref{eq:integral_equation2}) into a
fourth order ordinary differential equation. A proof for the existence
of a single-pulse of (\ref{eq:integral_equation2}) then becomes a
proof for the existence of a homoclinic orbit of the ODE. Since the
ODE has discontinuities across the threshold points, the ODE on the
real line is reduced to three different linear ODEs on three regions
separated by threshold points.  The matching conditions for the
solutions of the ODEs across the threshold points must satisfy a
system of five equations. From this 
system, we are able to construct different single-pulse solutions.

The eigenvalue structure of the linear ODEs is important for
determining how many pulses exist. For real $\omega_1$ and $\omega_2$,
there are at most two pulses, the small single-pulse and the
large one. Amari's case ($\alpha=0$) belongs to this regime. The large
single-pulse can transform to a dimple-pulse depending on the
threshold value (Fig.~\ref{fig:realPbump}.) If the eigenvalues are
complex, there could be a small single-pulse and two
large pulses with different widths. Depending on the threshold, these two
large pulses could be dimple-pulses (Fig.~\ref{fig:comp1dimple}.)
There also exists a transition point where a dimple-pulse breaks into
a double-pulse. 

There are three ways that the large pulse can
disappear. First, for fixed gain $\alpha$ and threshold $u_{\scriptscriptstyle T},$
if the excitation is too strong, {\it i.e.} ratio ${\displaystyle A/a}$ is large,
the width of the large pulse becomes wider
and eventually loses existence.  
Second, with fixed excitation,
$i.e.,$ fixed $a$ and $A,$ if the gain is too large, $i.e.,$ $\alpha$ is
large, the large pulse increases in height and blows up at a finite
value of $\alpha$.  Third, with too little excitation or gain, 
the stable large pulse coalesces with the unstable small pulse and
vanishes in
a saddle node bifurcation.

The pulses are a proposed mechanism of persistent neuronal activity
observed during working memory. 
Therefore, it is crucial to access their stability.  In the
accompanying paper, we show that the large pulse is stable and the
small pulse is unstable.  We also show that dimple-pulses can be
stable.  We show that single-pulses can exist for a wide variety of
gain and connection functions.   However, for single-pulses to exist with
low firing rates we require the gain to not be too large and
inhibition to dominate excitation.  This suggests that the cortex
could be dominated by inhibition.

\bigskip

\noindent{\bf Acknowledgment:}
We would like to thank G.~Bard Ermentrout, Jonathan Rubin, Bjorn
Sandstede and William Troy for illuminating discussions.  This work
was supported by the National Institute of Mental Health and the
A.P. Sloan foundation. 


\nocite{Aliprantis0,Aliprantis1,Arbib,Atkinson,Bender,Boyce,Champneys1,Delevs,Duffy,Ellias,Enander,Ermentrout1,Ermentrout2,Folland,Fuster1,Garvan,Green,Griffel,Guo3,Gutkin,Haskell,Kato,Kuznetsov,Miller,Morrison,Murray,Nicholls,Nishiura,Pelinovsky,Polianin,Powers,Rahman,Rubin,Rubin1,Rudin,Salinas,Strogatz,Troy2,Troybook,Wiggins,Wilson1,XPPAUT,Zeidler} 

\bibliographystyle{plain}
\bibliography{paper1}

\end{document}